\documentclass{aa}
\usepackage{graphicx}
\usepackage{natbib}
\bibpunct{(}{)}{;}{a}{}{,}
\usepackage[varg]{txfonts}
\usepackage{siunitx}
\usepackage{tabularx}
\usepackage[export]{adjustbox}
\usepackage{placeins}

\begin{document}

\title{Classification of radio sources through self-supervised learning}

\author{
Nicolas Baron Perez \inst{\ref{inst1}}\and
Marcus Br{\"u}ggen\inst{\ref{inst1}}\and
Gregor Kasieczka\inst{\ref{inst2}}\and
Luisa Lucie-Smith\inst{\ref{inst1}}}

\institute{
Hamburger Sternwarte, Universit{\"a}t Hamburg, Gojenbergsweg 112, 21029 Hamburg, Germany\label{inst1}\and
Institut f{\"u}r Experimentalphysik, Universit{\"a}t Hamburg, Luruper Chaussee 149, 22761 Hamburg, Germany\label{inst2}
}

\date{Received date /
Accepted date }

\abstract
{The morphology of radio galaxies is indicative of their interaction with their surroundings, among other effects. Since modern radio surveys contain a large number of radio sources that would be impossible to analyse and classify manually, it is important to develop automatic schemes. Unlike other fields, which benefit from established theoretical frameworks and simulations, there are no such comprehensive models  built for
radio galaxies. This stands as a challenge to data analysis in this field and novel approaches are required.}
{In this study, we investigate the classification of radio galaxies from the LOFAR Two-meter Sky Survey Data Release 2 (LoTSS-DR2) using self-supervised learning.}
{Our deep clustering classification strategy involves three main steps: (i) self-supervised pre-training; (ii) fine-tuning using a labelled subsample created from the learned representations; and (iii) performing a final classification of the selected unlabelled sample. To enhance morphological information in the representations, we developed an additional random augmentation, called a random structural view  (RSV).}
{Our results demonstrate that the learned representations contain rich morphological information, enabling the creation of a labelled subsample that effectively captures the morphological diversity within the unlabelled sample. Additionally, the classification of the unlabelled sample into 12 morphological classes yields robust class probabilities.}
{We successfully demonstrated that a subset of radio galaxies from LoTSS-DR2, encompassing diverse morphologies, can be classified using deep clustering based on self-supervised learning. The methodology developed here bridges the gap left by the absence of simulations and theoretical models, offering a framework that can readily be applied to astronomical image analyses in other bands.}


\maketitle

\section{Introduction}

Deep, wide-area radio surveys are mapping a rapidly increasing number of radio sources. Recent and forthcoming surveys, including  LOFAR Two-Metre Sky Survey \citep[LoTSS;][]{2017A&A...598A.104S} and Evolutionary Map of the Universe \citep[EMU;][]{2011PASA...28..215N}, along with surveys utilising  Square Kilometre Array (SKA), are expected to identify millions of galaxies.

The large number and heterogeneity of radio galaxies makes them difficult to classify and historical schemes may no longer be adequate as they are no longer aligned with our physical understanding \citep{2021Galax...9...85R}. 

One particularly challenging aspect is the morphological classification that solely rests on the geometric appearance of the source. Morphological classification is very important  to our understanding of the physical processes that form radio galaxies, such as how the jets that are produced by the supermassive black hole are interacting (and have interacted in the past) with their environment. Moreover, it is important for studying the relation between radio galaxies and galaxy evolution and for assessing the energetics of the sources.

The visual categorisation of these radio sources is becoming increasingly time-intensive and will soon be impractical due to the rapidly expanding volumes of data. To address this challenge, citizen science projects, such as  Radio Galaxy Zoo \citep[RGZ;][]{2015MNRAS.453.2326B, 2025MNRAS.536.3488W}, have been deployed for the classification of astronomical sources.\\

Radio galaxies can present compact or extended radio morphologies \citep{2017MNRAS.466.4346M} and are traditionally classified into Fanaroff-Riley (FR) classes, either FRI (core-bright) or FRII (edge-bright) galaxies \citep{fr}. There are several differences between the two classes: FRIs show less powerful jets than FRIIs, and these jets are disrupted quite close to the core of the radio galaxy while jets in FRII radio galaxies stay relativistic for much larger distances. In addition to the FRI/FRII classification, radio galaxies have been classified as doubles, double-doubles, triples, narrow-angle tails (NATs), wide-angle tails (WATs), bent-tails, hybrid, X-shaped, S-shaped, C-shaped, core-dominant, one-sided sources, and others \citep{2004rcfg.proc..335K}. Hybrid sources have the potential to reveal the origin of different radio morphologies \citep{2022ApJ...941..136S}.
There is no unified scheme and \cite{2021Galax...9...85R} has advocated for the tagging of radio galaxies instead. This was followed up using a natural language processing algorithm to derive a radio galaxy morphology taxonomy \citep{2023MNRAS.522.2584B}. 

With the advent of LOFAR, this classification scheme and the resulting luminosity correlation have been revisited using a dataset that is larger by two orders of magnitude \citep{mingo}. This work found that for this larger sample, the source luminosity could not predict its morphological class, finding a strong overlap of both classes in luminosity. Moreover, different morphologies such as restarting or double-double FRIIs and hybrid sources have also been found \citep{mingo}.\\

With the excellent performance of convolutional neural networks (CNN) in classifying natural images, there have been several efforts to train CNNs to classify images of radio sources \cite[e.g.][]{2021MNRAS.505.1464M, 2023A&C....4400728L}. \cite{2019MNRAS.482.1211W} developed a radio source morphology classifier based on a region-based CNN that was trained with FIRST and WISE data using the classification scheme of number of components and flux peaks. More recently, \cite{2025ApJS..276...46L} found thousands of bent-tail radio galaxies in the FIRST survey using deep learning and visual inspection.

\cite{2019MNRAS.487.1729L} explored the performance of capsule network architectures against simpler CNN architectures, to classify unresolved, FRI, and FRII morphologies. \cite{2023MNRAS.522..292B} have further adapted CNNs for radio galaxy classification by (i) using principal component analysis (PCA) during pre-processing and (ii) by guiding the CNN to search for specific features within the image data. They found that this adaptation led to a more stable training process and could reduce overfitting. We also refer to \cite{2023NewAR..9701685N} for a recent review on the machine learning-aided classification of radio sources.\\

Given all these developments, a drawback of supervised classification that is particularly pertinent to astronomy is the relatively small number of labelled training data and the morphological incompleteness when compared to comprehensive surveys. \cite{wgan} explored the augmentation of labelled data using a conditional Wasserstein Generative Adversarial Network (wGAN) to improve the classification accuracy of compact, FRI, FRII, and bent sources.

Another drawback is the varying classification schemes that have emerged in recent years, driven by the continuous nature of the diverse morphology of radio galaxies, which can lead to inconsistencies and confusion. Therefore, \cite{2021Galax...9...85R} proposes the use of tags instead of discontinuous boxes. The advantage of using tags is the possibility to combine them and extend the list of tags depending on the observed sources.\\

In recent years, self-supervised learning (SSL) has demonstrated remarkable results, particularly in the field of computer vision classification, the relevant area for this work - for an overview of recent developments in SSL, refer to \cite{ssl_cookbook}; for specific applications in astrophysics, see \cite{ssl_astro}. This training strategy allows models to learn from large amounts of unlabelled data by extracting meaningful representations from the data's inherent structures. Consequently, this approach overcomes the limitations of explicitly labelled datasets, resulting in a more robust model due to its ability to process and learn from a large volume of data.

Subsequently, the model can be `fine-tuned' by continuing its training with a more carefully curated and labelled dataset, using the weights obtained from the `pre-training' phase. This fine-tuning phase helps to adapt the general knowledge that the model learned during pre-training to specific `downstream tasks'. These are particular tasks that the model is designed to assist with, such as image classification. The combination of SSL pre-training followed by fine-tuning is commonly referred to in the literature as semi-supervised learning.

Three distinct types of SSL methods have emerged: invariance-based methods, generative methods, and what we term `context-aware' methods. The first type employs hand-crafted data augmentations to learn representations that remain invariant under these image transformations. The second type involves corrupting or removing parts of images and training the model to predict these missing sections. The third type focuses on predicting representations for specific parts of an image based on representations from other parts. For a more comprehensive discussion on these categories of SSL and the presentation of the first context-aware method, we direct readers to the work by \cite{2023arXiv230108243A}.

Efforts to use semi-supervised learning techniques in astrophysics include gamma-ray blazar classification \citep{2024MNRAS.528..976B}, cosmological inference \citep{2024MNRAS.527.7459A}, semi-supervised classification of FRI and FRII radio galaxies \citep{2023PrCS..222..601H}, hierarchical fine-tuning of a convolutional autoencoder \citep[AE;][]{2006Sci...313..504H} to classify six types of radio galaxies \citep{2019ApJS..240...34M}, a comprehensive study on semi-supervised classification of radio galaxies \citep{2022MNRAS.514.2599S}, and similarity search for optical galaxies \citep{2021arXiv211013151S}. Moreover, \cite{2024RASTI...3...19S} showed that a self-supervised model can be used for high-accuracy classifications and similarity searches of radio sources. Recently, \cite{2024arXiv241114078C} conducted a benchmark study to evaluate the performance of various invariance-based SSL algorithms on radio astronomy datasets employing different classification schemes. As part of this research, a new labelled dataset was developed using data from multiple radio surveys.

SSL algorithms learn a mapping from image space into an abstract vector space, distributing the dataset according to the features identified by the encoder. If the learned representation captures morphological information -- which is heavily influenced by the learning task -- the way images are distributed within this vector space should reflect the continuous nature of radio galaxy morphology. This enables a shift away from the discontinuous classification boxes on a preliminary level, while offreing the ability to search for higher density regions that contain similar objects at a higher level.

The unsupervised classification was  performed for multiple astronomical datasets. Works based on clustering algorithms include \cite{2024ApJ...962..164T} for high-redshift galaxies, \cite{2024MNRAS.528.4852P} for X-ray objects, \cite{2024A&A...683A.104L} for molecular clumps, and \cite{2022A&A...663A..21D} for galaxy spectra.

Unsupervised classification using deep clustering, which can incorporate the combination of SSL with clustering algorithms \citep{10585323}, has been applied to galaxy surveys in the near-infrared band \citep{2024ApJ...961...51V}, in the thermal infrared \citep{2022MNRAS.517.1837G} and in the visual band \citep{2021ApJ...921..177S}. In the context of radio observations, \cite{2024MNRAS.tmp..951M} classified FRI and FRII radio galaxies with this technique.

Moreover, the unsupervised classification of radio sources exceeding FRI and FRII galaxies with a self-organising map \citep[SOM;][]{2001som..book.....K, 2024A&C....4700824V} was studied first in combination with an autoencoder \citep{SANGER1989459} and the clustering algorithm $k$-means \citep{lloyd1982least} by \cite{2019PASP..131j8011R}. Later, the unsupervised classification with a SOM and followed manual grouping of classes was performed within \cite{2021A&A...645A..89M}. \\

In this work, we investigate the classification of radio sources from the LoTSS Data Release 2 \citep[LoTSS-DR2;][]{lotss_dr2} using deep clustering based on an invariance-based SSL method. We also introduce a new data augmentation, namely, the random structural view (RSV). The dataset used for pre-training consists of an size and image-quality filtered subset of the entire release, which we refer to as the `unlabelled sample', and it includes significant morphological diversity. In our approach, we forgo the use of previously labelled datasets to mitigate morphological biases stemming from label incompleteness. Instead, we generate a labelled subsample using the learned representations customised to the morphological diversity of the unlabelled sample. We tackle this classification task with a three-step process: (i) model pre-training, wherein we learn morphologically informed representations; (ii) model fine-tuning, in which we adjust the model for the classification task; and (iii) classification downstream task, where we classify the unlabelled sample.

This paper is structured as follows. Section~\ref{sect:data} provides a detailed description of the data used in this work. Section~\ref{sect:pretraining} outlines the methodology for pre-training the model and presents the corresponding results. In Sect.~\ref{sect:finetuning}, we discuss the fine-tuning process of the model and its associated outcomes. Section~\ref{sect:results} highlights the classification results of the unlabelled subsample, serving as our downstream task. In Sect.~\ref{sect:discussion}, we interpret our findings and put them in context with existing literature. Finally, in Sect.~\ref{sect:conclusions}, we summarise our conclusions and insights drawn from the study.

\section{Data}
\label{sect:data}

This section outlines the data utilised in our study. The real data served multiple purposes: it was employed for pre-training the model and constructing the tailored labelled subsample used for fine-tuning. It represented the unlabelled sample targeted for classification in our downstream task. Meanwhile, the simulated data was used for model evaluation during the pre-training phase.

\subsection{LOFAR DR2 data}

In this work, we used the synthesised image products from LoTSS-DR2 with a resolution of $\SI{6}{\arcsec}$. This data release comprises observations covering $27\%$ of the northern sky within the frequency range of $120-\SI{168}{\mega\hertz}$, for more details, we refer the reader to \cite{lotss_dr2}.

We used the publicly available\footnote{\url{https://lofar-surveys.org/dr2_release.html}} mosaics and the radio-optical cross-matched catalogue \citep{catalogue}. The full catalogue initially contained $\num{4116934}$ radio sources. We discarded the unresolved sources since they are generally simple to classify. Additionally, we excluded the sources that presented errors during image production or where the final image contains \texttt{NaN} values (the image production process is described in detail below). Moreover, we removed sources with largest angular sizes (LAS) smaller than $\SI{30}{\arcsec}$ since we focus on morphological classification of resolved sources, which span more than five beam sizes. This selection led to $\num{253242}$ sources. We then applied filters that aid in training the model for this pilot study, resulting in our unlabelled sample, listed below.
\begin{itemize}
    \item We restricted the dataset to sources with available optical coordinates, which resolved the ambiguity introduced when using radio mean positions. For instance, with radio mean positions, images of centre-bright sources are centred at the galaxy core. Meanwhile, the images of edge-bright sources with strongly asymmetric hotspot intensity were centred at the most luminous hotspot. This reduces the number of sources to $\num{192593}$.
    \item We applied a filter on the peak flux, selecting sources with $F_{\text{peak}} > \SI{0.75}{mJy/beam}$.
    We were left with $\num{76235}$ sources.
    \item We limited the LAS range to $\SI{30}{\arcsec}<{\rm LAS}<\SI{60}{\arcsec}$ to reduce the variety of morphologies within a certain established radio galaxy morphology due to source resolution. For instance, a centre-bright radio galaxy with low angular resolution can be described by the linear arrangement of a circularly shaped bright core and two elliptically shaped lobes. Conversely, with a higher angular resolution, the image will show a better-resolved core with jets departing from a perfect linear shape and lobes that present greater details. We observed that including all sources in $\textrm{LAS} > \SI{30}{\arcsec}$ leads to considerably worse results. After this cut, we maintained $\num{43769}$ sources.
    \item We set a threshold of 300 pixels on the number of active pixels present in a circular region that fills the entire image. We defined an active pixel as a pixel with a value $v_{\rm pix} > 0.2$. This led to a final dataset of $\num{42230}$ sources.
\end{itemize}

The unlabelled sample comprises 16.7\% of all LoTSS-DR2 resolved sources that extend beyond five beam sizes. In this study, we filtered the data quite crudely, leading to substantial reductions. A more refined data selection strategy could potentially increase the fraction of classified radio sources.

In the following, we describe the production of the images fed into the network. First, we determine the mosaic with its centre closest to the radio source, using the haversine distance to account for mosaic overlaps. Subsequently, we extract a square region centred on the optical position of the host. The cutout has a side length of $s_{L} = 1.5 \times \textrm{LAS}$.  Subsequently, the pixel values below $v_{\rm min} = 3 \times \sigma_{\rm img}$ are clipped to $v_{\rm min}$, where $\sigma_{\rm img}$ is the $\sigma$-clipped standard deviation of the cutout. This helps to suppress noise and enhance the signal-to-noise ratio by preventing the influence of low-intensity artefacts. Given the varying sizes of the cutouts, we resized the images to a common side length of $s_L = 128$ pixels. Finally, using the min-max normalisation, we rescaled the image pixels to values of $v_{\rm pix}\in [0, 1]$.

We present several randomly selected examples in Fig.~\ref{fig:example_images}. The images illustrate that the morphology of the radio galaxy remains exceptionally complex after filtering.

\subsection{Simulated dataset}

The morphological complexity of the real dataset motivates the creation of a synthetic dataset based on a very simplified radio galaxy model. This dataset can be used to evaluate the model's basic learning process (see Sect.~\ref{sect:ssl_metrics}).

In this model, we represent a radio galaxy as a source with a core and two lobes. These three components are represented using two-dimensional Gaussian intensity distributions, with the core modelled by a circular Gaussian and the lobes by elliptical Gaussians. We consider two morphological degrees of freedom: (i) centre or edge-bright and (ii) linear or bent-shaped. The former is achieved by adapting the relative intensity of the core and lobes (an invisible core is also included), while the latter is implemented with an angle between the two lobes (where the core is the vertex of the angle). Hence, our simplified synthetic dataset consists of four morphological classes with source bending standing out as a crucial morphological characteristic.

\begin{figure}
\begin{center}
\includegraphics[width=0.99\columnwidth]{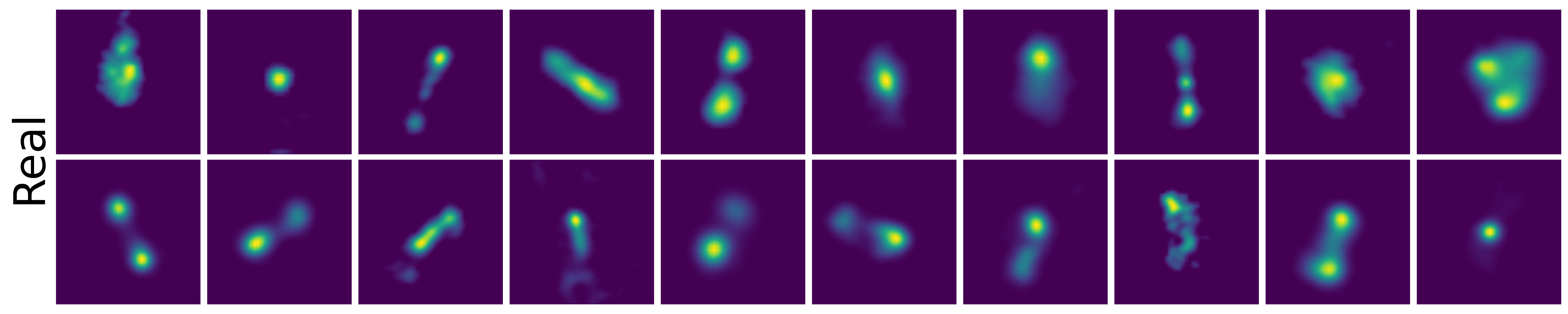}
\includegraphics[width=0.99\columnwidth]{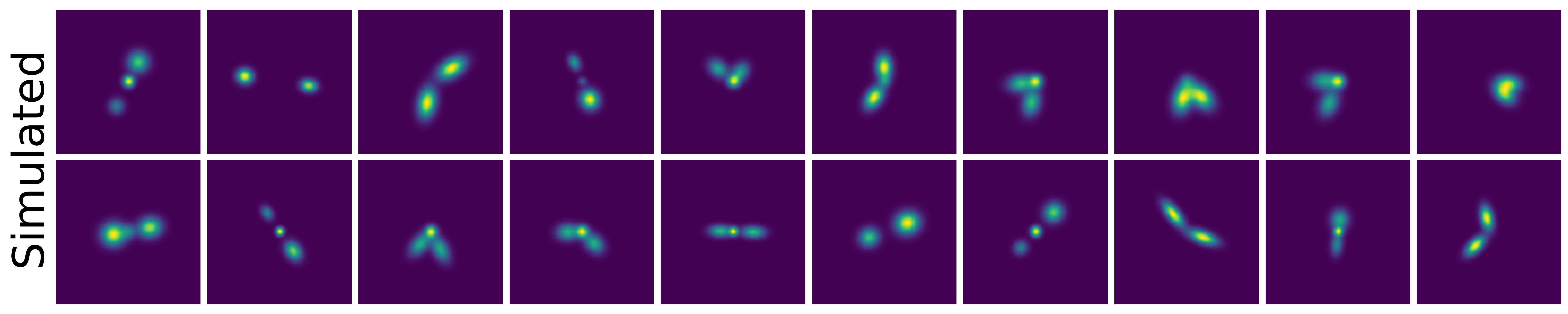}
\end{center}
\caption{Randomly selected images from both the real and simulated datasets, showcasing their morphological diversity.}
\label{fig:example_images}
\end{figure}

The linearly shaped mock sources take into account the following morphological parameters: source size, orientation angle, the four standard deviations of the lobe Gaussian distributions, and if it is centre or edge-bright. Finally, the bent-shaped mock sources have the parameters: orientation angle, lobe angle, two standard deviations for both identical lobes, and centre or edge-bright. Most of the mock source parameters were sampled from uniform distributions. We generated \num{5000} examples of each of the four morphological classes to have a balanced dataset. Randomly selected examples are shown in Fig.~\ref{fig:example_images}.

\section{Model pre-training}
\label{sect:pretraining}

This section details the model pre-training process, which uses invariance-based SSL techniques with the unlabelled sample. We describe our model architecture and training strategy, introduce our novel data augmentation approach, discuss the various methods used to evaluate the pre-training performance, and present the results of the pre-training stage.

\begin{figure*}
\begin{center}
\includegraphics[width=0.9\textwidth]{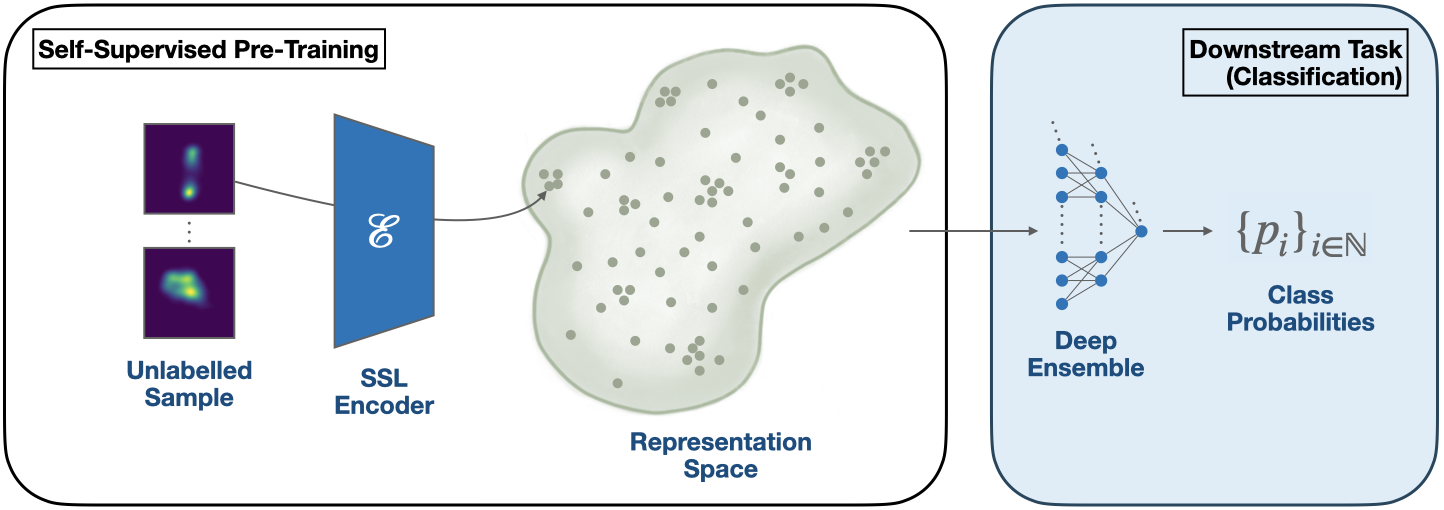}
\end{center}
\caption{Diagram illustrating the classification process. Initially, the encoder maps the input sample into the representation space, where calculations such as evaluating image similarities and determining distances to morphological classes can be performed. Subsequently, the deep ensemble generates robust class probabilities based on these representations.}
\label{fig:pipeline_inference}
\end{figure*}

Self-supervised algorithms are trained by addressing a broader task that is the pretext task, rather than focusing on predicting a specific value (the human-annotated label) as in supervised learning. The goal of SSL is to generate useful representations of the data without labels. Here, the representation space should group sources together based on morphology. These representations can then be used for downstream tasks. The classification process, which is the downstream task we focus on in this study, is illustrated in Fig.~\ref{fig:pipeline_inference}.

\subsection{Self-supervised training}
\label{sect:ssl_training}

The invariance-based SSL task involves generating two augmented views of the same input image and training the model to produce similar representations for these views compared to representations of different images. SimCLR is among the initial SSL algorithms that demonstrated exceptional performance in the linear classification of its extracted representations; namely, the linear evaluation protocol \citep{simclr}. We opted to use SimCLR for representation extraction due to its simplicity. Other models tested included BYOL \citep{2020arXiv200607733G}, MSF \citep{2021arXiv210507269A}, ProPos \citep{2021arXiv211111821H}, and NNCLR \citep{2021arXiv210414548D}. The simplicity was prioritised at this stage to focus on developing the pipeline, with the goal of creating a classification scheme that comprehensively encompasses the morphological diversity of the dataset.

\begin{figure}
\begin{center}
\includegraphics[width=\columnwidth]{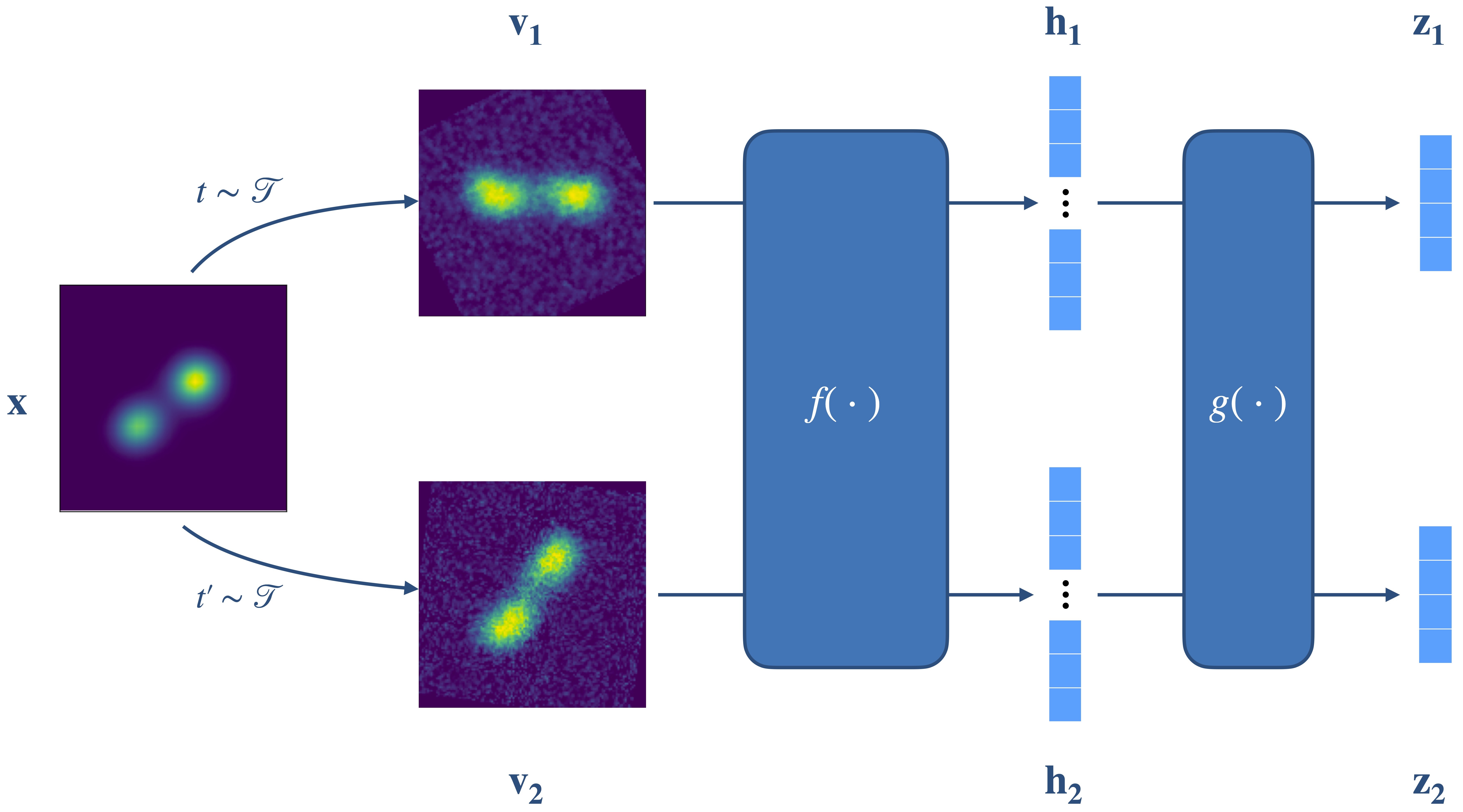}
\end{center}
\caption{SimCLR architecture: The input image $\mathbf{x}$ undergoes random transformation twice, generating two views. These views are then processed through the encoder $f(\cdot)$ to extract corresponding representations. Subsequently, an MLP $g(\cdot)$ is employed to project these representations into a lower dimensional space. The objective of the learning task is to maximise the similarity between these projections.}
\label{fig:simclr}
\end{figure}

In terms of its network architecture, SimCLR consists of an encoder $f(\cdot)$ responsible for extracting a representation vector, $\mathbf{h}$, from its input, $\mathbf{v}$, followed by a projector, $g(\cdot),$ that maps the representation, $\mathbf{h}$, into a projection $\mathbf{z}$ in a lower dimensional space (see Fig.~\ref{fig:simclr}). The input images have dimensions of 128x128 pixels, the representation consists of 512 dimensions, and the projection vectors have 256 dimensions. We employed a ResNet18 \citep{resnet} for our encoder $f(\cdot)$, as done in previous works \citep[e.g.][]{2024RASTI...3...19S, 2024MNRAS.tmp..951M}, and a multi-layer perceptron (MLP) with one hidden layer for the projector $g(\cdot)$.

To define the learning task of SimCLR, we generated two views, $\mathbf{v_1}, \mathbf{v_2}$, of each input image, $\mathbf{x}$, using random transformations (more details are given below). Thus, given a batch of size, $N$, we pass the $2N$ views through the networks, obtaining $2N$ projections. Then, the learning task is to ensure that projections $\mathbf{z_1}$ and $\mathbf{z_2}$ are closer in the learned space than $\mathbf{z_1}$ is to any of the remaining $2N-2$ projections. Mathematically, this is formulated by minimising the  normalised temperature-scaled cross-entropy (NT-Xent) loss function. For a pair of views, $\mathbf{v_i}, \mathbf{v_j}$, is given by:

\begin{equation}
\label{eq:ntxent_loss}
    \mathcal{L}^{\rm NT-Xent}_{\,i,j} = -\log\frac{\exp(\,\textrm{sim}(\bf{z}_i,\bf{z}_j)\,/\,\tau\,)}{\sum_{k=1}^{2N}I_{[k\neq i]}\,\exp(\,\textrm{sim}(\bf{z}_i,\bf{z}_k)\,/\,\tau\,) } \,,
\end{equation}
where $\textrm{sim}(\mathbf{u},\mathbf{v}) = \mathbf{u}^\intercal\mathbf{v}\,/\,||\mathbf{u}||\,||\mathbf{v}||$ represents the cosine similarity, $\tau$ is a temperature parameter, and $I_{[k\neq i]}$ is 1 if $k\neq i$ and 0 otherwise. The temperature parameter adjusts the influence of the data points within the batch: a smaller value gives greater weight to closer representations, while a larger value increases the influence of representations that are further apart \citep{2019arXiv190201889F}.

A key component of SimCLR is the set of random transformations used to generate image views that define the learning task. We employ standard augmentations: random vertical flips with $50\%$ probability, random rotations between $-360^\circ$ and $360^\circ$, random brightness jitter with a factor between 0 and 1.1, and random resize crop with a scale between 0.9 and 1. Additionally, we incorporate a custom augmentation, which is detailed in Sect.~\ref{sect:random_augmentation}.

We trained the SimCLR model to minimise the NT-Xent loss function with a learning rate (LR) of $0.0003$ and a batch size of 1024. The LR schedule starts with a linear warm-up \citep{2017arXiv170602677G} phase of 5 epochs starting at half the LR and is succeeded by a cosine annealing \citep{2016arXiv160803983L} phase of 200 epochs that finishes at 1/50 of the LR. We used AdamW \citep{2017arXiv171105101L} as the optimiser with a weight decay of $0.0001$ and a temperature of $\tau=0.09$ in the NT-Xent loss. The temperature value is chosen empirically. It should be small enough to account for low-level morphological properties, yet large enough to ensure that representations of sources with the same morphology are not separated in the representation space due to any non-morphological properties. For downstream tasks, we normalised the representations predicted with the encoder to obtain an Euclidean norm of 1, as done when computing the NT-Xent loss function during training.

\subsection{Random structural view}
\label{sect:random_augmentation}

So far, the two image views used to compute the NT-Xent loss were produced from the same input image. These are used as positive pairs, while all other $2N-2$ views in the batch are considered negative examples. For our classification purpose, this set-up is not ideal because sources with the same morphology, which appear identical, are treated as negative examples. Furthermore, sources with different morphologies, such as three-component FRI and FRII sources, can share numerous characteristics, with the only difference being the relative brightness of the core and the lobe hotspots. However, the random augmentations considered so far cannot differentiate between the sources based on this information.

To solve this issue, we used a metric to measure the similarity between images, based on the assumption that images of sources with identical morphology will have a higher similarity value compared to those with different morphologies. Referring back to the previous example, the similarity score between two three-component FRI will be higher than that between a three-component FRI and an FRII. Therefore, for a given source, we can select one of the sources with the highest similarity scores as a positive example. This allows us to create a random augmentation that can be used with a certain probability to replace one augmented view of the original source with an augmented view of a `very similar' source. We refer to this random augmentation as the RSV.

The similarity metric we use is the structural similarity index measure (SSIM), which was designed to imitate the human visual system under the assumption that the human eye is very sensitive to changes in structural information \citep{2004ITIP...13..600W}. Moreover, we use the following procedure to make the metric semi-invariant (SI-SSIM): For a pair of images $(\mathbf{x_i},\mathbf{x_j})$, SI-SSIM is defined as the maximum value of SSIM between $\mathbf{x_i}$ and eight transformed versions of $\mathbf{x_j}$. The transformations considered are: the original image, rotations by $90^\circ$, $180^\circ$, and $270^\circ$; a vertical flip; and a vertical flip combined with each of the aforementioned rotations.

To optimise the computation time, we compute the SI-SSIM matrix $M^{\rm img}(\mathbf{x_i},\mathbf{x_j})$ before starting the training of the SSL algorithm for a given dataset. For the computation, we make use of GPU power to accelerate the calculation. Depending of the dataset size, we opted to reduce the image resolution to further accelerate the SI-SSIM calculation.

During the training, for a given image $\mathbf{x_i}$, we draw one image from the $k=4$ most similar images in the batch (where $k=0$ refers to the image $\mathbf{x_i}$ itself) and substitute the second augmented view of $\mathbf{x_i}$ with the second augmented view of the drawn image. This approach yields a $(k-1)/k$ probability for the  RSV. The intended choice of a small $k$ is to ensure that the most similar images present in the batch have almost identical morphology. We provide an overview of all random augmentations used during pre-training and their respective parameters in Table \ref{tab:random_augmentations}.

We present an ablation study of this random augmentation in Appendix \ref{sect:ablation}. The results indicate that the RSV enhances morphological clustering in the representation space and facilitates the identification of additional, more complex morphological classes, particularly bent sources. Limitations of the random augmentation and suggestions for future testing are discussed in Sect.~\ref{sect:discussion}.

\begin{table}
\caption{Parameters of random augmentations.}
\label{tab:random_augmentations}
\centering
{
\renewcommand{\arraystretch}{1.1}
\begin{tabularx}{\columnwidth}{l l c}
\hline
\textbf{Random augmentation} & \textbf{Parameter} & \textbf{Value} \\
\hline
\rule{0pt}{2ex}
random vertical flip & probability & 0.5 \\
\rule{0pt}{3ex}
random rotation & probability & 1.0 \\
 & range of degrees & (-360, 360) \\
\rule{0pt}{3ex}
random brightness jitter & probability & 1.0 \\
 & contrast factor & (0, 1.1) \\
\rule{0pt}{3ex}
random resized crop & probability & 1.0 \\
 & scale & (0.9, 1) \\
\rule{0pt}{3ex}
random structural view & probability & 0.75 \\
 & neighbours & 4
\end{tabularx}
}
\end{table}

\subsection{Dimensionality reduction and clustering of representations}
\label{sect:clustering}

To examine the morphological information encapsulated in the representations, the most direct approach is to employ clustering algorithms. However, given the extensive processing time required for clustering algorithms on high-dimensional data, we applied PCA to reduce the dimensionality of the 512-dimensional vectors while preserving $95\%$ of the total variance. This reduction results in 22 principal components, minimising the component correlation and organising them according to the variance they explain. We then applied a clustering algorithm to identify high-density regions within the transformed principal component space.

We tested the clustering algorithms $k$-means, HDBSCAN, BIRD, and agglomerative clustering and found that Hierarchical Density-Based Spatial Clustering of Applications with Noise\footnote{\url{https://hdbscan.readthedocs.io/en/latest/}} \cite[HDBSCAN;][]{hdbscan} worked best for our application. This clustering algorithm can handle clusters of different densities and accounts for `noise', namely, data points that lie far from any dense region that defines the clusters. Since we are working with uncurated data, our dataset will include outliers, corrupted data, and noise. It can also detect non-spherical clusters, which is important since we do not have any reason to assume that the emerging high-density regions in the principal component space should be spherical. 

HDBSCAN includes mainly two parameters to adjust the clustering, the parameter \texttt{min\_samples}, which determines how conservative the clustering should be, and the minimal cluster size \texttt{min\_cluster\_size}. Setting $\texttt{min\_samples}=1$ works best for our data.

The clustering analysis involves an iterative application of HDBSCAN on the PCA-processed representations to determine the \texttt{min\_cluster\_size} that yields the highest number of clusters. We found that $\texttt{min\_cluster\_size}=19$ produced the highest number of clusters, totalling 95. Our objective in identifying the maximum cluster count is to obtain a set of sources within the clusters that captures the morphological diversity of the unlabelled sample as completely as possible.

\subsection{SSL evaluation metrics}
\label{sect:ssl_metrics}
In the following, we present a short description of the metrics used to evaluate the learning performance of the pre-trained model. Before outlining the specific evaluation metrics, we introduce the Soft Nearest Neighbour Loss \citep[SNNL;][]{pmlr-v2-salakhutdinov07a, 2019arXiv190201889F}, which we used as an evaluation metric during pre-training. SNNL measures the relative distances in the representation space of a given image by comparing the distances between this image and others with the same label in the batch to the distances between this image and all images in the batch. For a more detailed explanation of SNNL, we refer to Sect.~\ref{sect:adaptation}.
The evaluation metrics utilised during pre-training are as follows.
\begin{itemize}
    \item Top-$k$ accuracy: This metric measures the percentage of input data where the cosine similarity between the projections of both corresponding augmented views ranks among the top $k$ in the batch.
    \item Mean position accuracy: After computing the cosine similarity for all $2N$ projections in a batch, this metric provides the average rank of the cosine similarity between the augmented view projections for each input image, averaged over all $N$ images in the batch.
    \item SNN metric: SNNL was employed to assess how closely neighbouring images, whose distances are calculated using SI-SSIM in image space, are located in the representation space. We used a temperature parameter of $\tau=0.07$ and considered $k^{\rm SNN}=10$ SI-SSIM neighbours, which are used as labels for the metric computation.
    \item Linear evaluation protocol: This metric consists of the supervised classification of the representations obtained from the self-supervised encoder using a linear model. We use the simulated dataset to calculate the metric.
    \item Clustering: To understand the information encoded in the representations, we perform a clustering. The visual inspection of the resulting clusters provides insights into the overall structure of the representation space by contrasting different clusters, while examining the morphological homogeneity within each cluster reveals the finer details of the structure.
\end{itemize}

Our goal is to obtain representations that primarily capture morphological information, ensuring that sources with the same morphology have nearly indistinguishable representations. Thus, instead of focusing solely on maximising top-$k$ accuracy or minimising mean position accuracy, we aim to improve these metrics to a point of equilibrium, enabled by the high representation similarity among sources within the same class.

Moreover, an improvement in the SNN metric indicates that the fine-grained distribution of the representation space effectively captures morphological similarity. While the improvement of the linear evaluation protocol denotes that the representations encode the notion of centre or edge-bright and linear or bent-shaped. Ultimately, the evaluation of the clustering provides the most significant insights into the information encapsulated within the representations.

\subsection{Pre-training results}

We examine the representations derived from the pre-training phase and the clusters that have been identified. We begin by evaluating the training metrics. The NT-Xent starts at 6.79 and decreases steadily throughout training, eventually reaching 4.48, indicating effective learning. Post-training, we achieve a top-1 accuracy of $18.9\%$, a top-5 accuracy of $32.7\%$ and a mean position accuracy of 49. Although the accuracy values are not exceptionally high, they show consistent improvement over the course of training. This trend is expected and desirable, as our primary goal is not to distinguish individual sources but to develop clusters of morphologically similar sources that are nearly indistinguishable from each other.

\begin{figure}
\begin{center}
\includegraphics[width=0.99\columnwidth]{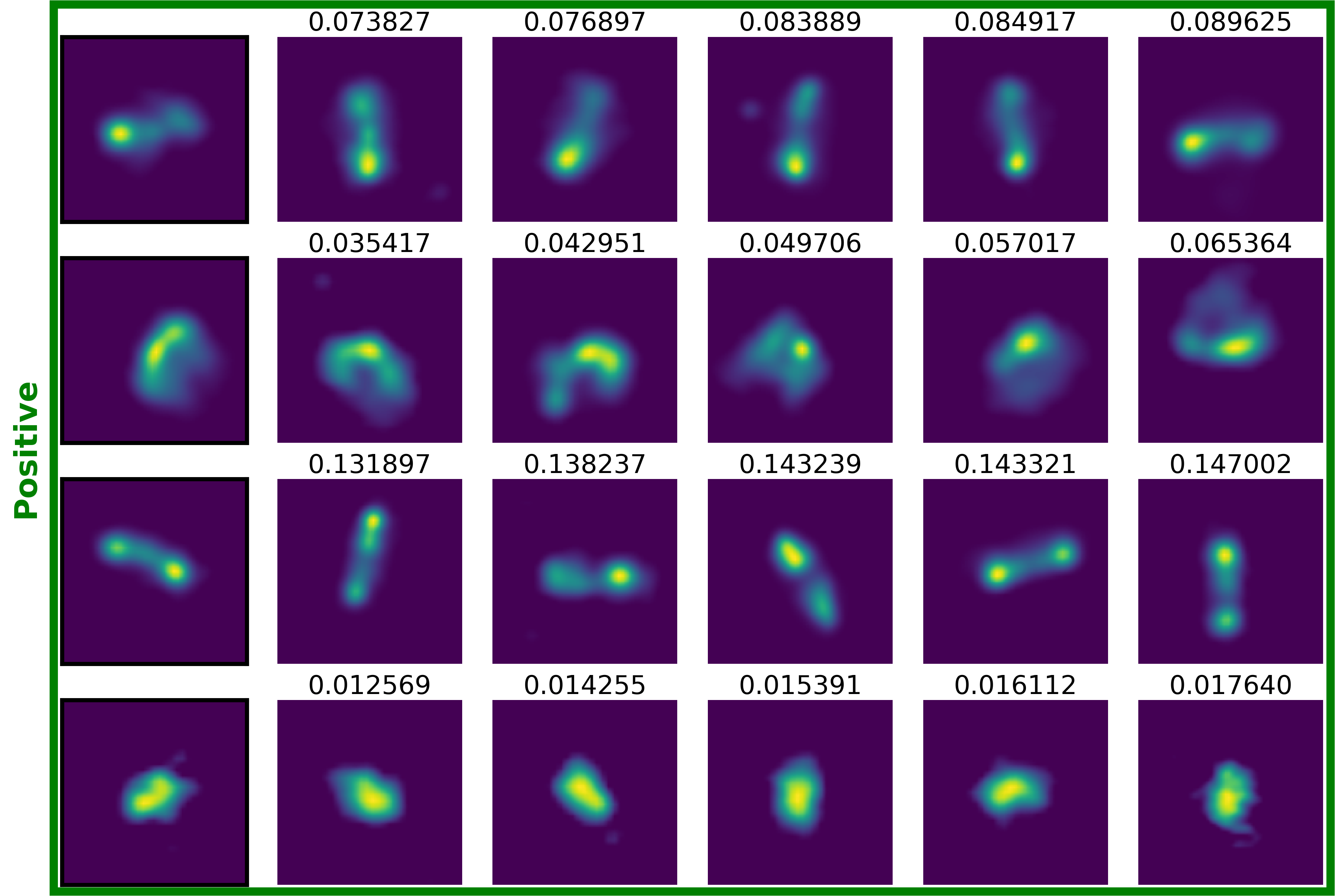}
\includegraphics[width=0.99\columnwidth]{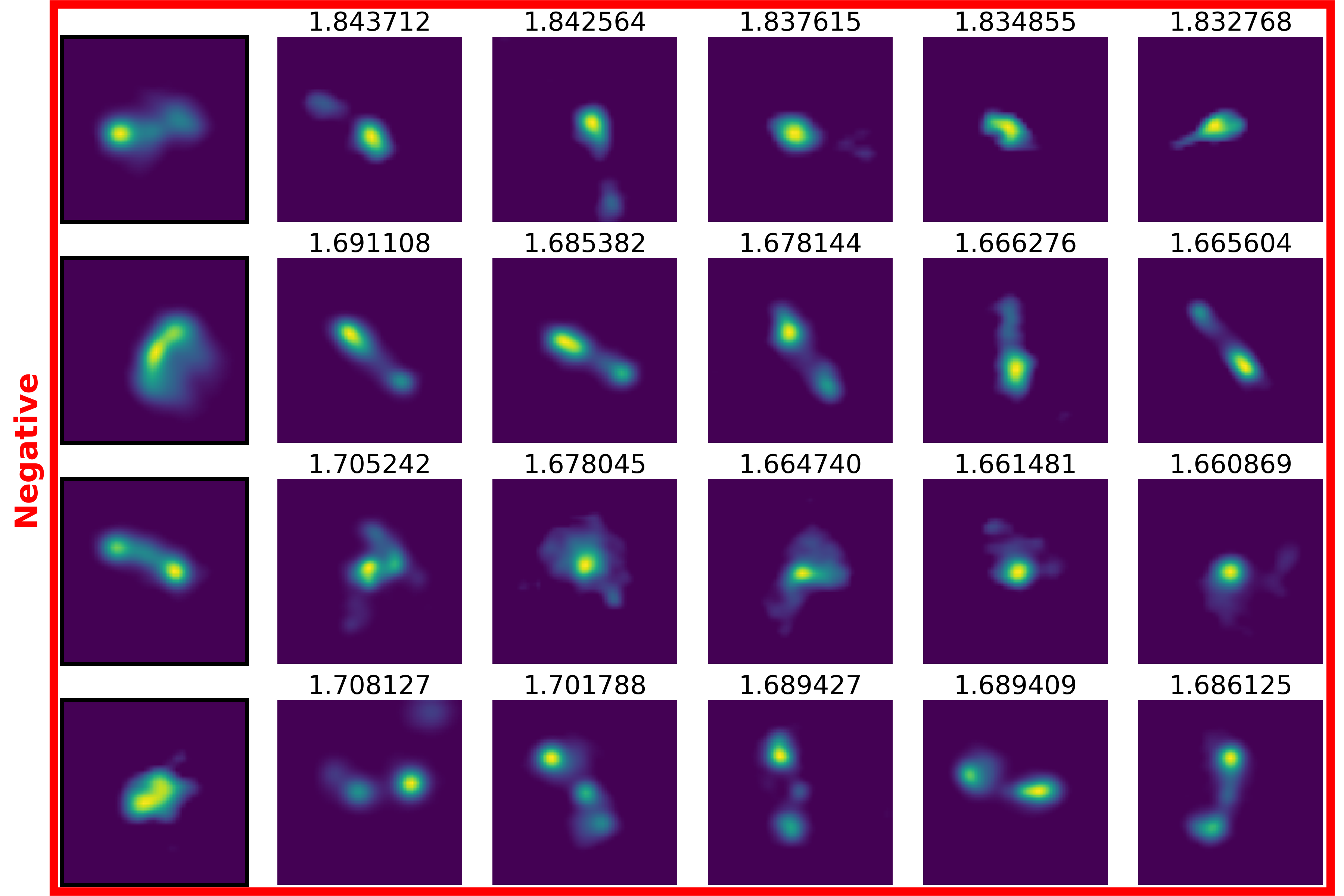}
\end{center}
\caption{ Query images, shown in the first column. The first four rows show the five nearest neighbours based on cosine distance, while the last four rows display the five farthest representations. Above each neighbour, the corresponding cosine distance value is shown.}
\label{fig:cosine_nn}
\end{figure}

The SNN metric value decreases from 3.47 to 1.68 during training, indicating that the ten most similar sources for each target source, based on the SI-SSIM, are becoming more closely aligned. This trend is evident when examining the images that are closest in representation space for any given source. To demonstrate this, we present the nearest and farthest neighbours of randomly selected images, determined by the cosine distance of the PCA-transformed representations, in Fig.~\ref{fig:cosine_nn}. Cosine distance is defined as ${\rm dist}(\mathbf{u},\mathbf{v}) = 1-{\rm sim}(\mathbf{u},\mathbf{v})$. This figure illustrates that the representations capture a wide range of ample morphological information within the PCA-transformed representation space. For closely situated representations, the images associated with them exhibit similar morphologies. Conversely, when comparing more distant representations, the corresponding images exhibit significantly different morphologies. Specially, morphological features, such as the bending angle, the relative intensity of radiation peaks, the number of source components, and source extension are effectively encoded within the representations.

\begin{figure}
\begin{center}
\includegraphics[width=0.9\columnwidth]{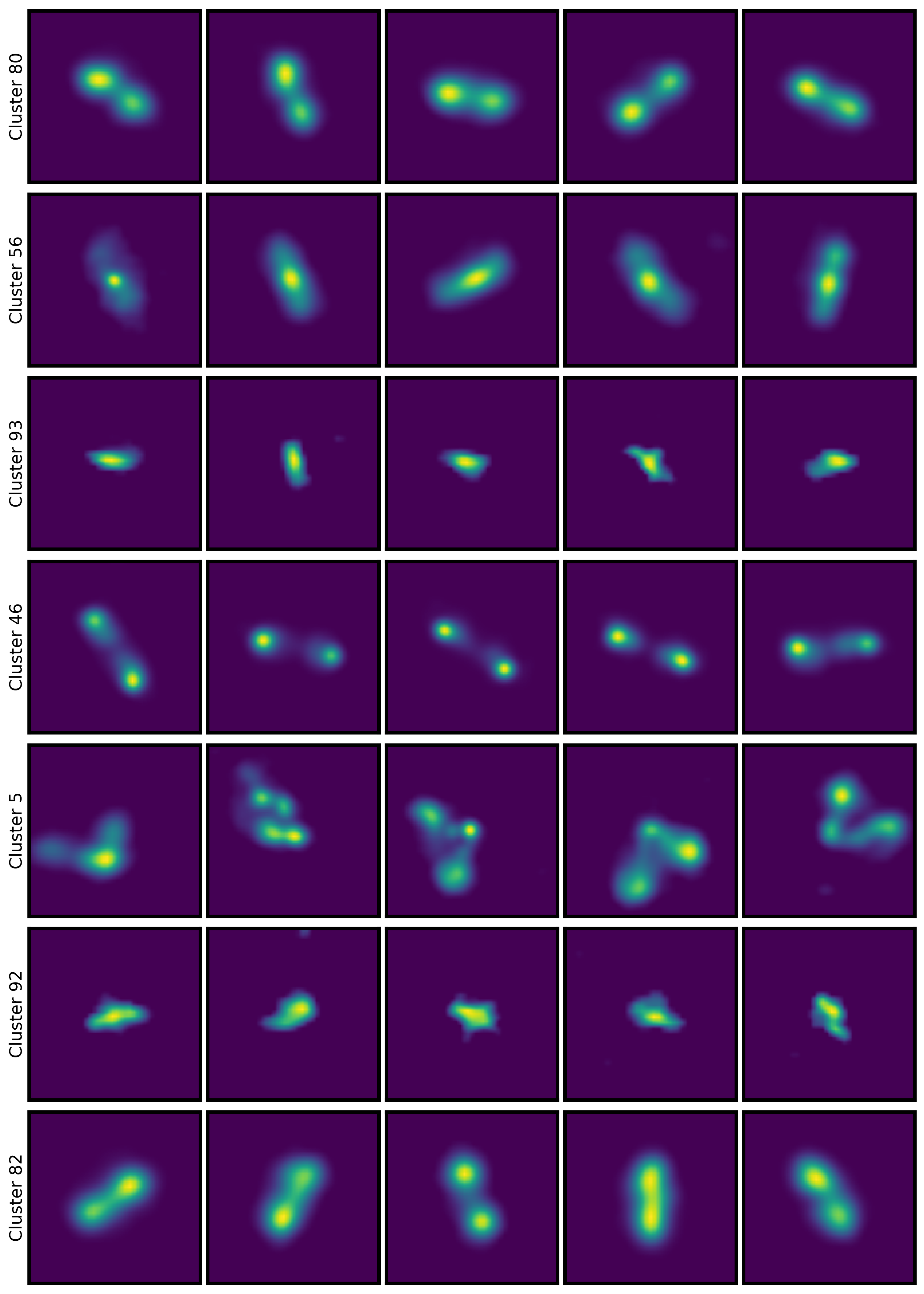}
\end{center}
\caption{Seven clusters were randomly selected and from each of these clusters, five members were also chosen at random to be displayed.}
\label{fig:random_cluster_members}
\end{figure}

Furthermore, we achieve linear evaluation accuracies of 90\%, 89\%, and 91\% for the simulated training, validation, and test datasets, respectively. The linear evaluation protocol involves classifying the representations using a model as simple as a linear one. Therefore, these results indicate that the encoder effectively distinguishes between linear and bent sources, as well as between centre-bright and edge-bright sources. This is evident in the local neighbourhood of any given query representation. To verify whether this morphological distinction extends to the 95 identified clusters, we present five randomly selected sources from seven randomly selected clusters in Fig.~\ref{fig:random_cluster_members}. The plot demonstrates that most of the clusters are homogeneous in terms of morphology. Specifically, the clusters shown contain sources with the following morphologies (from top to bottom): symmetric doubles, core-bright sources, amorphous sources, edge-bright sources with visible lobes, and bent sources. The last two clusters again feature amorphous and symmetric double sources, respectively.

\begin{figure}
\begin{center}
\includegraphics[width=0.99\columnwidth]{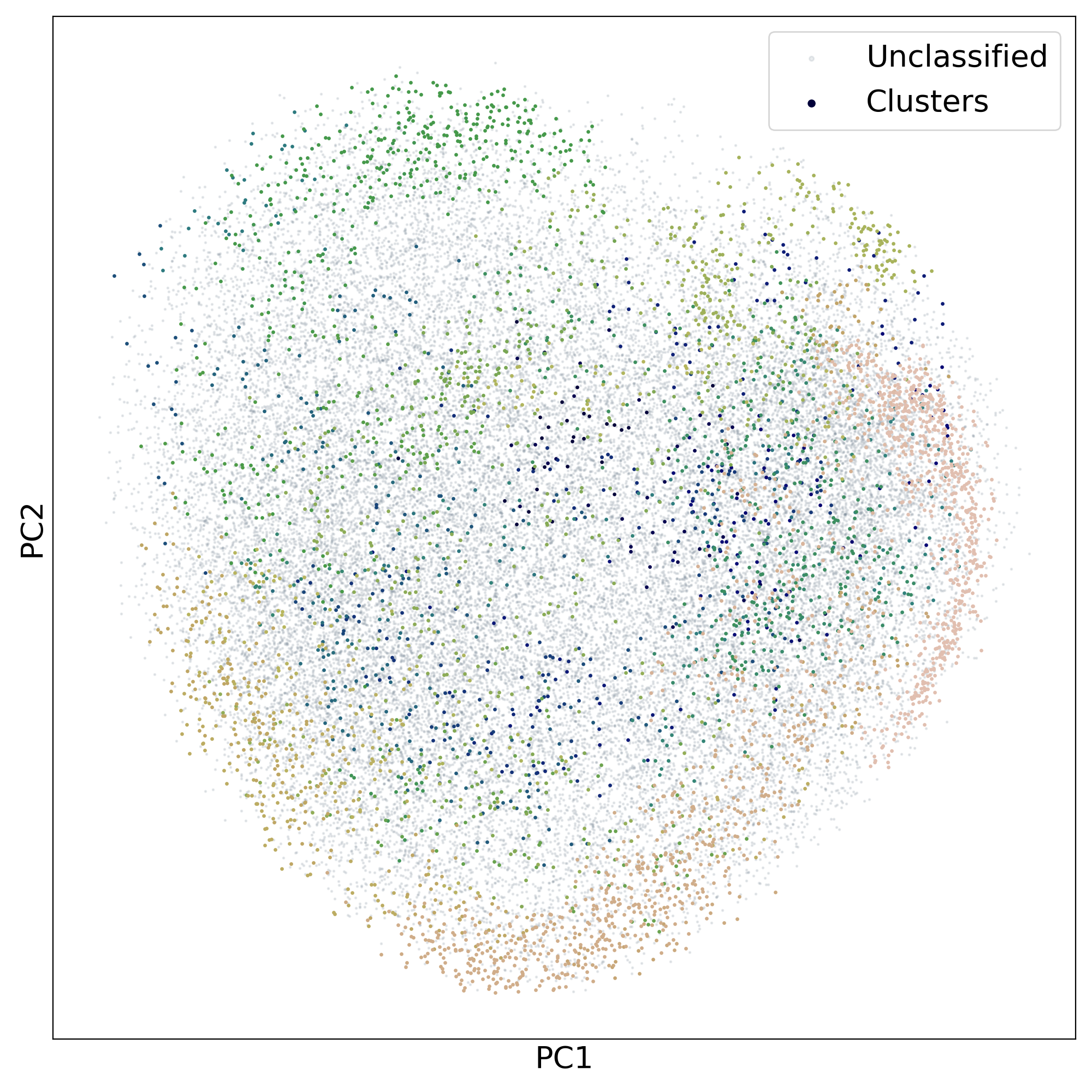}
\end{center}
\caption{Raw clustering results are depicted, illustrating the distribution of representations in the space defined by the first two principal components. Data points are colour-coded according to their cluster pre-labels with unclassified sources displayed in grey.}
\label{fig:clusters_ssl}
\end{figure}

We present the raw clustering results in Fig.~\ref{fig:clusters_ssl}. It shows the distribution of the first two principal components of the representations, with colours indicating cluster pre-labels and gray for unclassified sources. We consistently use the first two principal components throughout the paper to visualise the distribution of the representations in a two-dimensional space, which helps to emphasise the dominant structure of the data. Upon comparison with visualisations obtained using UMAP \citep{2018arXiv180203426M} and t-SNE \citep{vandermaaten08a}, no significant differences were observed. The plot reveals a diffuse structure within the PCA-transformed representations, making it challenging to visually distinguish individual clusters without colour differentiation. The clusters are dispersed across the entire representation distribution, indicating that they encompass a wide variety of morphological shapes.

\begin{figure}
\begin{center}
\includegraphics[width=0.99\columnwidth]{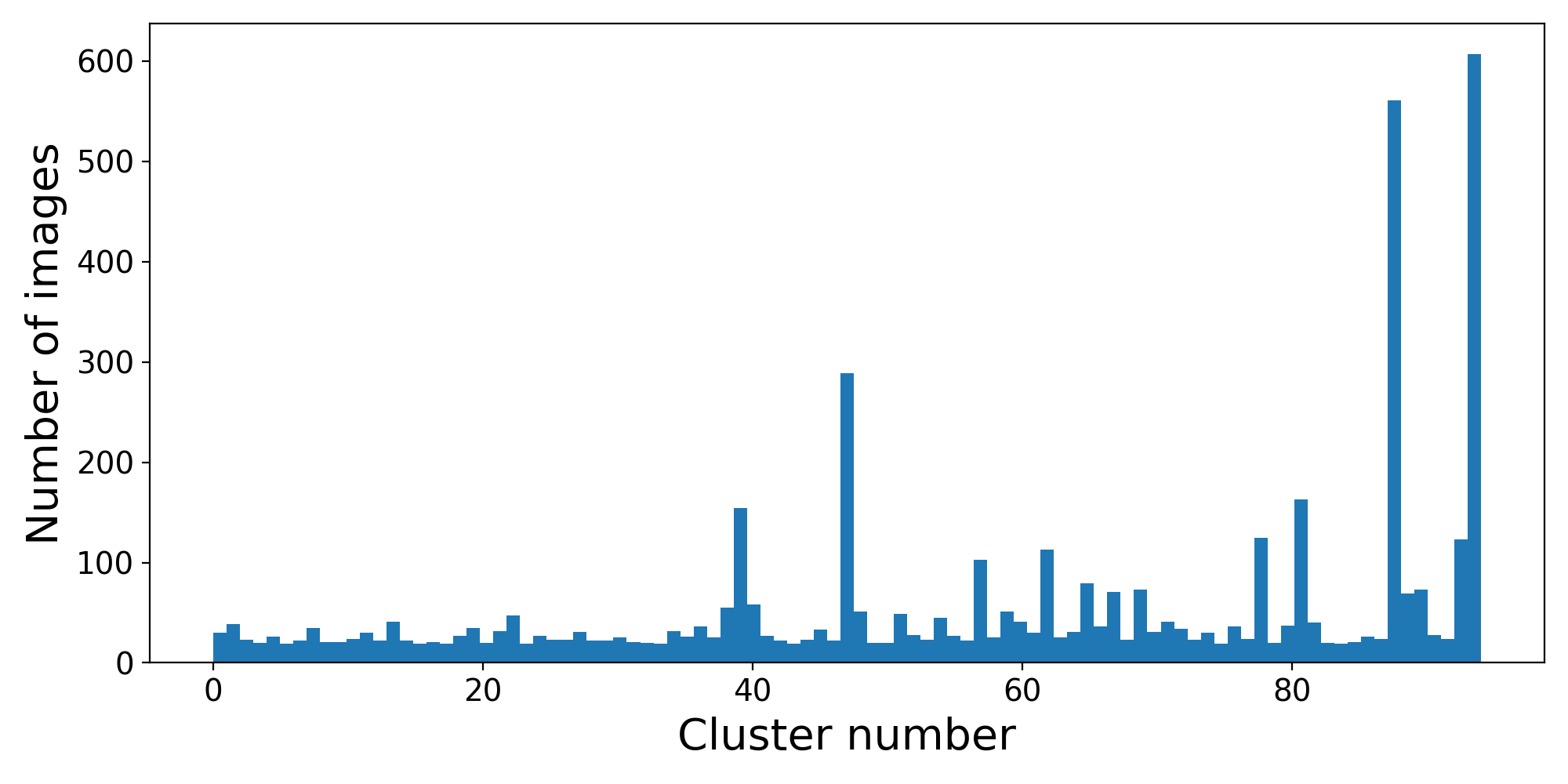}
\end{center}
\caption{Size distribution of clusters forming the pre-labels. At this stage, the cluster pre-labels are numerically assigned from 0 to 94.}
\label{fig:pre_cluster_distribution}
\end{figure}

Owing to the diffuse nature of the representation distribution (Fig.~\ref{fig:clusters_ssl}), we configured the algorithm to identify numerous small clusters that capture a low fraction of the dataset, but include clusters with homogenous source morphologies. The identified clusters encompass 11.6\% of the unlabelled sample and most of them have a small number of members, as shown in Fig.~\ref{fig:pre_cluster_distribution}, with an average cluster size of 51.5. Visual inspection reveals that 14 clusters exhibit mixed morphology, where, for example, the sources have common shapes but differ in intensity peak positions.

\section{Model fine-tuning}
\label{sect:finetuning}
In this section, we outline our model fine-tuning procedure, designed to adapt the network for improved performance in predicting cluster labels for each image in the unlabelled sample (our downstream task). We explain how the labelled subsample used for fine-tuning is constructed and describe the final composition of this subsample. Additionally, we introduce the fine-tuning procedure itself and present the results it yields. This procedure involves training a deep ensemble to achieve more accurate class probability predictions.

The necessity of fine-tuning arises from the SSL pre-training, which involves a broad learning task and results in representations that include information not particularly useful for the classification task. Therefore, the goal of fine-tuning is to adapt the representation space for improved classification performance. The fine-tuning task involves predicting the class probability of images with assigned cluster label. Simultaneously, the fine-tuned representation space naturally adapts to accommodate the continuous nature of the radio source morphologies.

\subsection{Assigning cluster labels to images}
\label{sect:cluster_labels}
Constructing the labelled subsample for the fine-tuning step involved pairing images with their assigned labels, where each label represents the cluster to which the image belongs. The clusters should comprehensively capture the full morphological diversity present in the unlabelled sample.

We started with assigning labels to images based on the unsupervised clustering strategy described in Sect.~\ref{sect:clustering}. The resulting cluster labels are termed `pre-labels' in the subsequent discussion. Using these cluster pre-labels as ground truth class labels for the supervised fine-tuned model would not return meaningful class predictions. In our approach, this is because the clustering method produces large numbers of redundant classes that also include imperfections, whereby morphologically similar images can be assigned to different clusters and divergent images could end up belonging to the same cluster.

Instead, our goal was to design a training set which includes only images with a high probability of belonging to their parent cluster, so as to avoid confusing the model with images that are poorly correlated to their parent cluster. Therefore, we restricted to the purest subset of (images, cluster label) pairs.

Contrary to standard supervised classification tasks, our setting does not suffer from the issue of potentially having a smaller number of training set images. This is thanks to the optimisation conducted during the SSL step using a large dataset. For further details on this characteristic of SSL pre-trained models in the context of radio source classification, we refer to \cite{2024RASTI...3...19S}.

The label assignment procedure involved two steps: (i) reducing the number of redundant clusters and (ii) assigning labels to the most homogeneous set of images within each cluster.

To minimise redundant clusters, we visually inspected random samples from each cluster. Clusters that were determined to contain sources with highly heterogenous morphologies had their pre-labels replaced with the `unclassified' label. Subsequently, clusters containing sources that appeared to have identical morphologies but different pre-labels were assigned a common label.

We then computed the SI-SSIM matrix for each remaining cluster and averaged over one dimension to obtain a mean SI-SSIM value for each cluster member. Based on the observed morphological homogeneity within each cluster, we retained members at the 90th, 80th, or 60th percentile according to their mean SI-SSIM values. The images falling below these thresholds were reclassified as unclassified, while the remaining images retained their respective cluster labels. In this way, we assigned labels to the most homogeneous set of images within each cluster.
In a future work, we plan to explore fully automated approaches for constructing training set labels from the representation space.

\subsection{Labelled subsample}
We present the labelled subsample used for fine-tuning, specifically designed to capture the morphological diversity inherent in the unlabelled sample targeted for classification in our downstream task. From the initial 95 cluster pre-labels, we relabelled 26 clusters as `unclassified' due to the heterogenous morphologies of the images within each cluster (as described in the first step of Sect.~\ref{sect:cluster_labels}). We identified three potential explanations for the formation of these clusters: source miscentering, images with poor quality, and sources with common shapes but different intensity peak positions.

Additionally, we consolidated the clusters containing sources with the same morphology, reducing the total number of cluster pre-labels from 69 to 12 (second step in Sect.~\ref{sect:cluster_labels}). We defined and described these source types as follows:
\begin{itemize}
    \item Artifact: mainly sources that are outshined by another source present in the cutout. Due to the other source higher luminosity, the source in question is not visible.
    \item Amorphous: sources with an unclear morphology and of a small size, although the image size is proportional to the source LAS provided in the catalogue.
    \item Bright core:  sources that are single Gaussian intensity dots. They might be compact radio sources, also called FR0s \citep{2023A&ARv..31....3B}.
    \item Head-tail: sources with a bright core and a single lobe connected to the core.
    \item Single lobe:  sources with a bright core and a single diffuse lobe separated from the core.
    \item Centre-bright: centre-bright sources with visible core and lobes.
    \item Centrally peaked ellipse: sources with elongated structure and an intensity peak in the central region, but no clear core and lobe structure.
    \item Symmetric double:  sources have low resolution and appear as two Gaussian intensity regions.
    \item Edge-bright:  sources present bright hotspots; for some, the core and the lobes are visible.
    \item Diffuse bent: sources with bent morphologies, appearing diffuse and including centre-bright and edge-bright structures.
    \item Structured bent: bent sources with well-defined radio structures, exhibiting mixed brightening (edge and centre).
    \item Circular diffuse:  sources have a relatively circular shape with intensity peaks inside the more diffuse emission. From a morphological point of view, they could be associated with radio halos. This could be confirmed by cross-matching with X-ray observations.
\end{itemize}

\begin{figure}
\begin{center}
\includegraphics[width=\columnwidth]{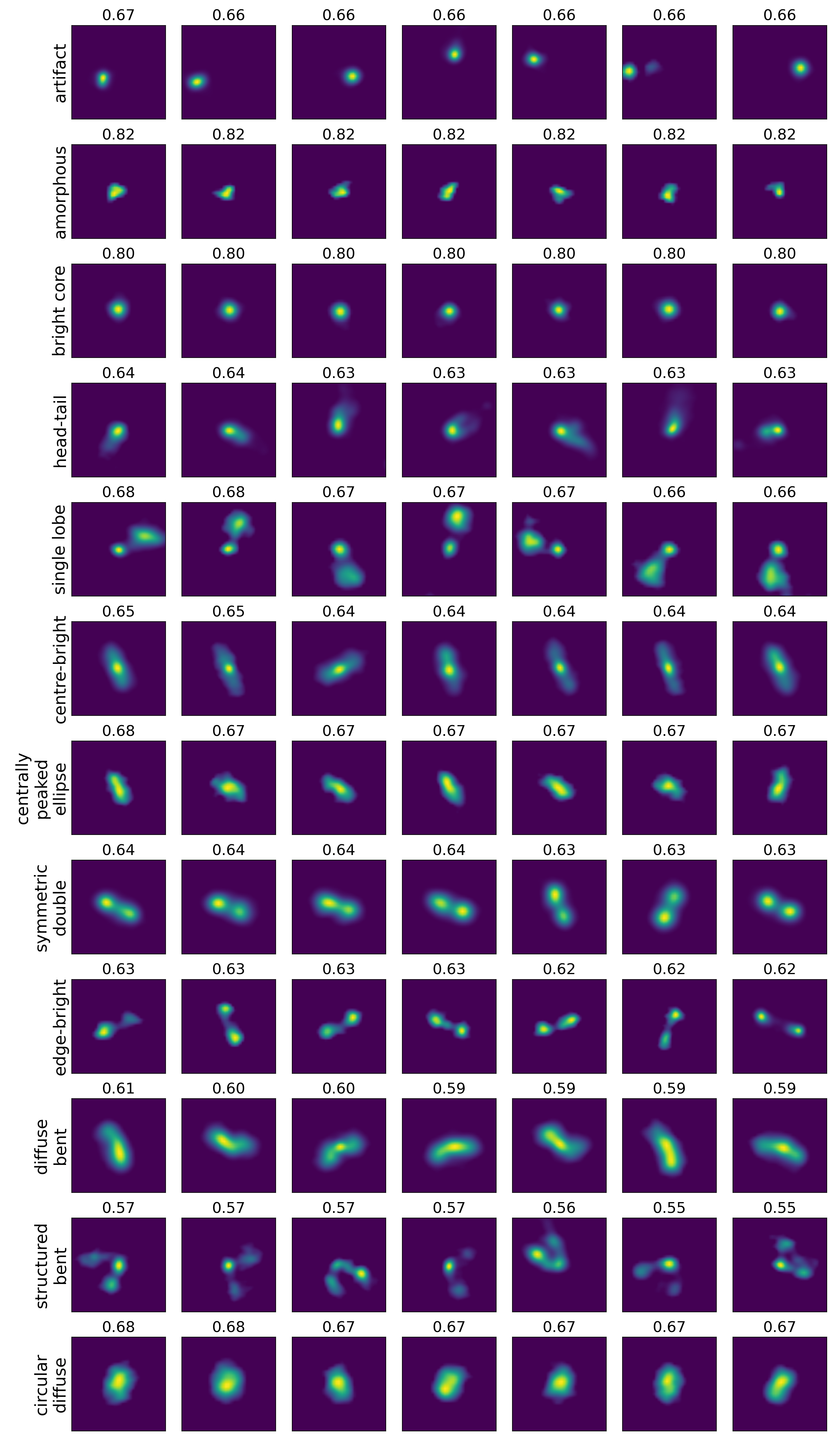}
\end{center}
\caption{Distinct clusters given in each row,   showcasing the seven sources with the highest mean SI-SSIM values to highlight the typical morphology exhibited by the cluster. The mean SI-SSIM value of each source is displayed above each panel.}
\label{fig:mean_ssim_clusters}
\end{figure}

The labelled subsample is highly imbalanced, with the largest cluster containing over 30 times as many sources as the smallest cluster. Figure~\ref{fig:mean_ssim_clusters} presents the members from each cluster with the highest mean SI-SSIM values, highlighting the typical morphology associated with each cluster.

\begin{figure*}
\begin{center}
\includegraphics[width=0.9\textwidth]{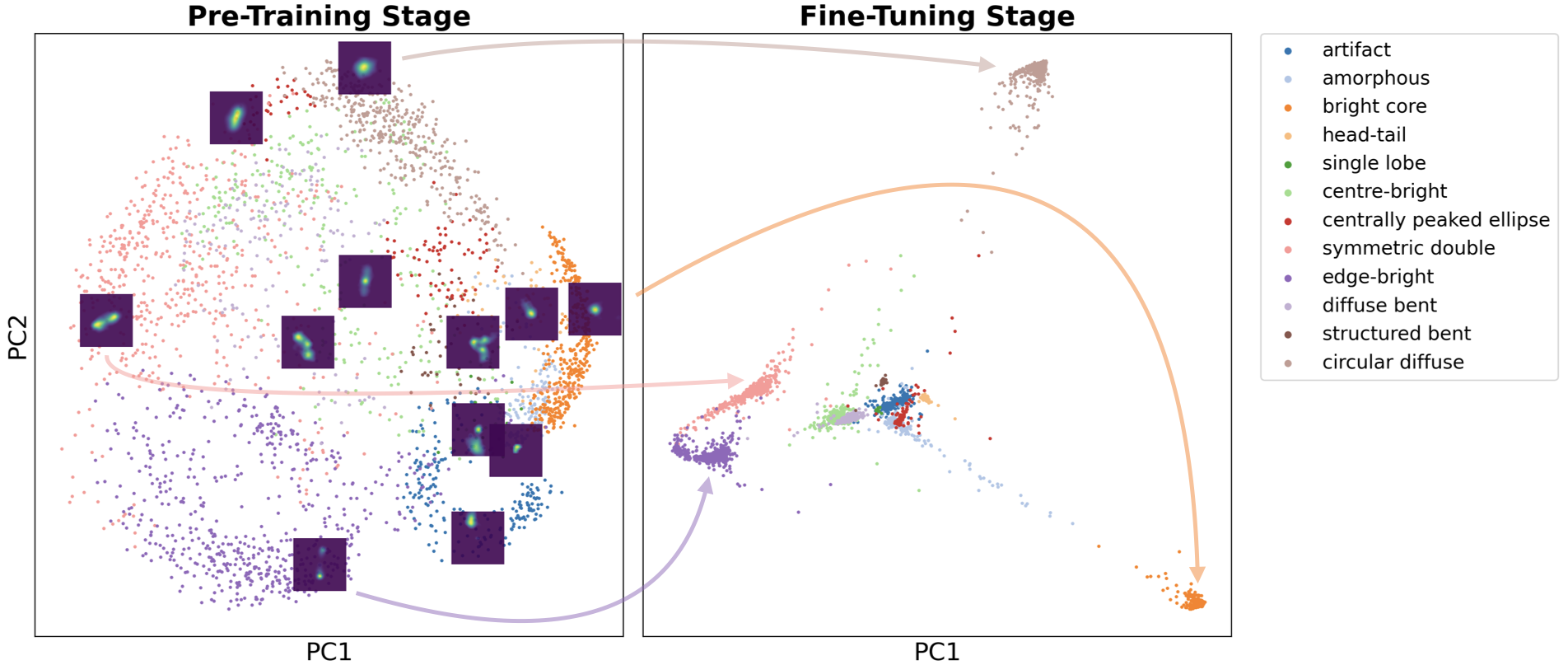}
\end{center}
\caption{Representations of sources from the labelled subsample within the space defined by the first two principal components, with colours corresponding to cluster labels as listed in the legend. \textbf{Left:} Representations from the pre-training stage are displayed, showing a dispersed distribution pattern. Additionally, a representative image from each cluster is provided to highlight the morphological structure within the distribution. \textbf{Right:} Post fine-tuning representations are presented, demonstrating compact clustering according to the labels. Arrows are included to depict the evolution of selected clusters from the pre-training to the fine-tuning stage.}
\label{fig:pretrain_finetune_evol}
\end{figure*}

Next, we examined the distribution of sources corresponding to each cluster label within the space defined by the first two principal components of the representations from the pre-training stage. This distribution is illustrated in the left plot of Fig.~\ref{fig:pretrain_finetune_evol}, where each cluster is distinguished by a unique colour. Additionally, a representative image is included to illustrate the typical morphology characteristic of that cluster. Most of the clusters are located within distinct, bounded regions, separated from others. However, some clusters exhibit significant overlap within this two-dimensional space. Qualitatively, we observed three trends that describe the representation space structure: 
\begin{itemize}
\item Considering only the clusters at the edge of the distribution, we find that single-component sources are located on the northeastern side, while double-component sources are on the southwestern side of the distribution. 

\item The fraction of active pixels grows anti-clockwise starting at the eastern part (bright core sources) and ending at the northern part (circular diffuse sources) of the representation distribution. 

\item The morphological complexity, reflected by the number of source components or bending angle, increases towards the centre of the distribution.
\end{itemize}

\subsection{Adapting the model to classification task}
\label{sect:adaptation}
With the constructed labelled subsample, we proceeded to fine-tune the encoder using the cluster labels by adding an MLP on top of the encoder. The supervised learning task employed for classification during this fine-tuning process allows the network to learn features that result in more meaningful representation encoding and more accurate class predictions than those obtained through self-supervision.

We used two weighted loss functions during fine-tuning, the cross-entropy loss, and SNNL. The loss weights accounted for the imbalanced nature of the dataset. The weight for class $i$ was calculated as $w_i = 1/n_i \times s_d/n_c$, where $n_i$ represents the number of sources in class, $i$, then $s_d$ is the dataset size, and $n_c$ is the number of classes.

The cross-entropy loss is a widely used function for multi-class classification tasks and is given as:

\begin{equation}
    \mathcal{L}^{\rm Xent} = -\sum_{c=1}^{n_c} y_c \log(p_c)\,,
\end{equation}
where $n_c$ is the number of classes, $y_c$ is 1 if the image belongs to class, $c,$ (and 0 otherwise), and $p_c$ is the predicted class probability.

SNNL was initially introduced in \cite{pmlr-v2-salakhutdinov07a} to fine-tune a metric learning algorithm and enhance the $k$-nearest neighbour classification. Subsequently, \cite{2019arXiv190201889F} improved this loss function and applied it to assess the degree of class entanglement in the representation space. A high value of SNNL indicates a significant overlap of the classes, while a low value implies that the classes are well separated and form compact clusters in the representation space \cite{2019arXiv190201889F}. The minimisation of SNNL decreases the ratio of the sum of distances between the member pairs of class, $i$, over the sum of distances between all data pairs in the batch. Mathematically this is given by:
\begin{equation}
\label{eq:snn_loss}
    \mathcal{L}^{\rm SNN}_i = - \log \frac{ \sum_{j \neq i, \, y_j = y_i,\,\, j=1,...,N} \,\exp(\,\textrm{sim}(\mathbf{z_i},\mathbf{z_j})/\tau^{\rm SNN}\,) }{ \sum_{k \neq i,\,\, k=1,...,N} \,\exp(\,\textrm{sim}(\mathbf{z_i},\mathbf{z_k})/\tau^{\rm SNN}\,) }\,,
\end{equation}
where $y_i$ is the class label of the i-th data point and $\tau^{\rm SNN}$ is the temperature of SNNL.

The total fine tuning loss functions is then given by:

\begin{equation}
    \mathcal{L}^{\rm FT} = \alpha\,\mathcal{L}^{\rm Xent} + (1-\alpha)\,\mathcal{L}^{\rm SNN}\,,
\end{equation}
where $\alpha$ is the loss function weighting factor.

We fine-tuned the encoder optimising the validation accuracy with a batch size of 512 and a learning rate of 0.001 for 100 epochs. We used an SNNL temperature parameter of $\tau^{\rm SNN}=0.2$ and a loss function weighting factor of $\alpha=0.5$. A higher temperature value enhances the compactness of each cluster, while increasing the distance between clusters, particularly for those that differ significantly based on the extracted features.

\subsection{Deep ensemble training}
\label{sect:ensemble}
We selected deep ensembles to obtain more reliable probability estimates for our class predictions. \cite{2016arXiv161201474L} investigated how the training of a deep ensemble should be conducted to obtain well-calibrated (predicted uncertainties and actual results observed in repeated experiments agree) and generalisable predicted probabilities. They concluded that training an ensemble of independent, randomly initialised networks with cross-entropy loss and averaging the predicted probabilities leads to the desired result. In this case, the predicted probabilities are given by
\begin{equation}
    p(y|\mathbf{x}) = \frac{1}{M}\sum^M_{m=1}p_{\theta_m}(y|\mathbf{x},\theta_m)\,,
\end{equation}
where $y$ is the class label, $\mathbf{x}$ is the image, $M$ is the number of networks that constitute the ensemble, and $\theta_m$ represents the parameters of the $m$-th network.

We constructed our deep ensemble by fine-tuning the encoder 50 times, with the weights of the classification head randomly initialised for each iteration. The ensemble representations are derived by averaging the representations predicted by each member of the ensemble.

\subsection{Fine-tuning results}

We go on to analyse the results of the fine-tuning process. Utilising the labelled subsample during fine-tuning allowed us to achieve classification accuracies of 99.2\%, 99.2\%, and 97.2\% for training, validation, and test datasets, respectively.

The distribution of the PCA-transformed representations post fine-tuning is depicted in the right panel of Fig.~\ref{fig:pretrain_finetune_evol}. Most clusters are situated near the centre of the space defined by the first two principal components, while four clusters are positioned farther away. The most distant clusters are the bright core sources and the circular diffuse sources. These classes exhibit the most distinct morphologies, as single-component objects. The other two clusters that are notably distant from the others are the symmetric double and edge-bright sources, which are located close to each other. Both of these have unique morphologies characterised by two nearly identical components.

We find that the sources are organised into more compact clusters (right panel of Fig.~\ref{fig:pretrain_finetune_evol}) compared to their distribution at the pre-training stage (left panel of Fig.~\ref{fig:pretrain_finetune_evol}), with inter-cluster distances more strongly influenced by the morphological similarity of the clusters. The improved separation of clusters in the fine-tuned space is an anticipated outcome of the supervised learning task. In general, cluster compactness and inter-cluster distance can be further adjusted by tuning the temperature parameter of the SNNL. Despite this, some clusters still overlap. A visual inspection shows that overlapping clusters share morphological properties such as hotspot distance, source size, and resolution.

\section{Results}
\label{sect:results}

In this section, we present the classification of all sources in the unlabelled sample, reflecting the predictions made with the deep ensemble.

The number of sources in each class is listed in Table~\ref{tab:class_population}. The table includes counts for the entire unlabelled sample as well as for sources with a class probability greater than 90\%. Our analysis reveals that the rank order of the classes differs between these subsets. While the most populous class, symmetric double, and the least populous class, single lobe, remain consistent; the intermediate classes shift primarily by one position. Notably, centrally peaked ellipse sources drop three positions, while bright core sources rise by two positions. In the subset with high confidence, double sources are the most prevalent, comprising 33.8\% of this set. Additionally, 14.9\% of the sources are edge-bright, while 10.6\% are centre-bright. Furthermore, 19.2\% of the sources exhibit a bending angle. Lastly, images classified as contamination, namely, artifact and amorphous sources, account for 5.8\% of the dataset.

\begin{table}
\caption{Distribution of sources across morphological classes.}
\label{tab:class_population}
\centering
{
\renewcommand{\arraystretch}{1.3}
\begin{tabularx}{\columnwidth}{l l l l}
\hline
 & \textbf{Class name} & \textbf{Whole sample} & $\mathbf{p\geq 90\%}$\\
\hline
1 & artifact & 1258 & 910 \\
2 & amorphous & 873 & 555 \\
3 & bright core & 1130 & 919 \\
4 & head-tail & 2441 & 1292 \\
5 & single lobe & 607 & 228 \\
6 & centre-bright & 4386 & 2682 \\
7 & centrally peaked ellipse & 1506 & 616 \\
8 & symmetric double & 12544 & 8564 \\
9 & edge-bright & 6314 & 3774 \\
10 & diffuse bent & 7255 & 3721 \\
11 & structured bent & 2491 & 1151 \\
12 & circular diffuse & 1425 & 939 \\
\hline
 & Total & 42230 & 25351 \\
\hline
\end{tabularx}
}
\tablefoot{Source counts are provided for the entire unlabelled sample and for the subset of sources with a class probability greater than $90\%$.}
\end{table}

\begin{figure}
\begin{center}
\includegraphics[width=\columnwidth]{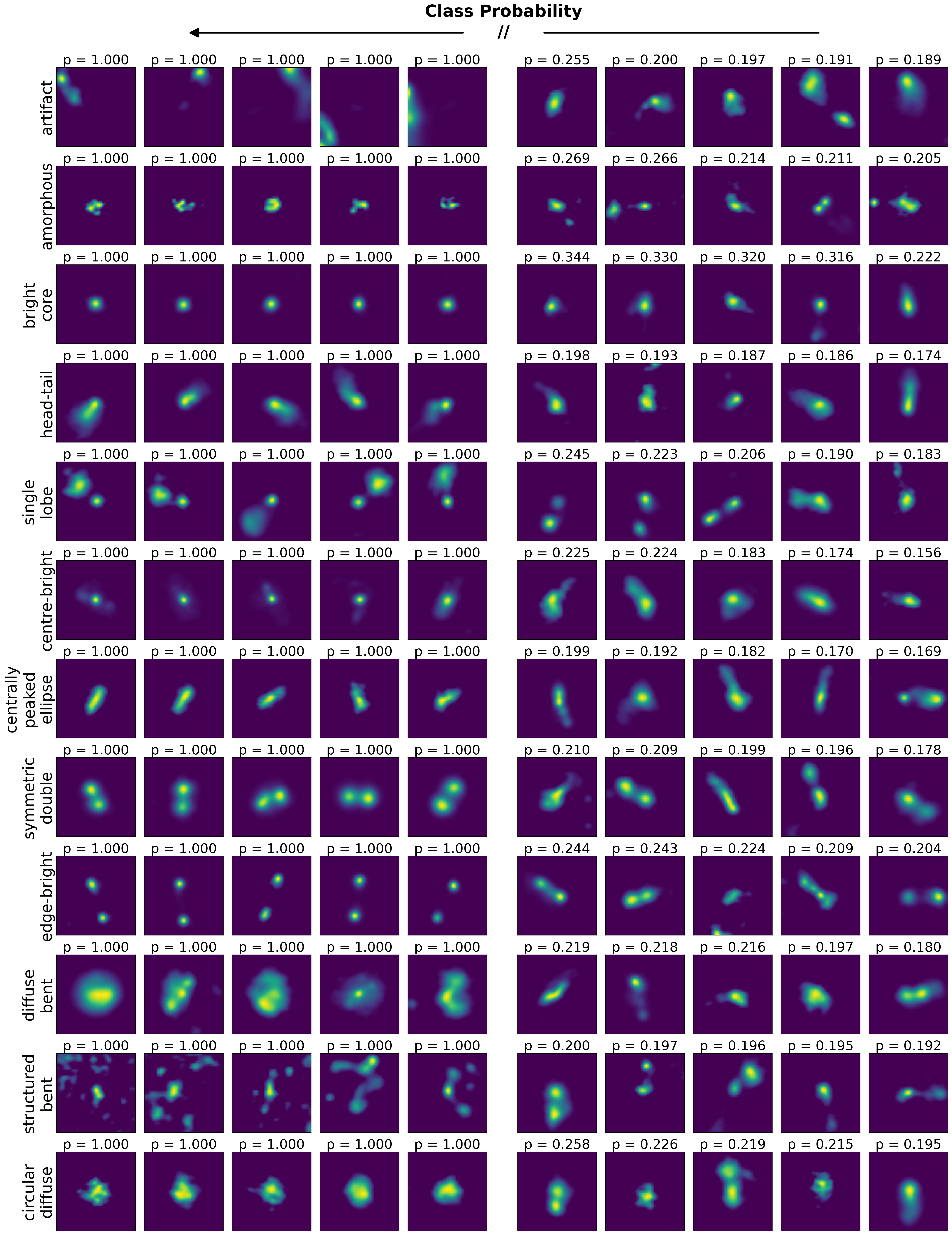}
\end{center}
\caption{Different classes shown row by row. The first five columns display the sources with the highest class probabilities, while the last five columns show those with the lowest class probabilities. This figure illustrates the model's confidence in its predictions. The class probability for each source is indicated at the top of each panel.}
\label{fig:final_class_members}
\end{figure}

In Fig.~\ref{fig:final_class_members}, we display the five sources with the highest and lowest class probabilities for each class. By showcasing these images, we illustrate the model's confidence in its predictions. The sources with the highest probabilities serve as prototypical examples that effectively represent most classes, particularly those with simpler morphology, helping to confirm the model's reliability in these categories. Meanwhile, examining the sources with the lowest probabilities helps identify ambiguous or challenging cases, highlighting areas where the model may struggle and providing insights into the potential areas for further improvement.

\begin{figure}
\begin{center}
\includegraphics[width=0.9\columnwidth]{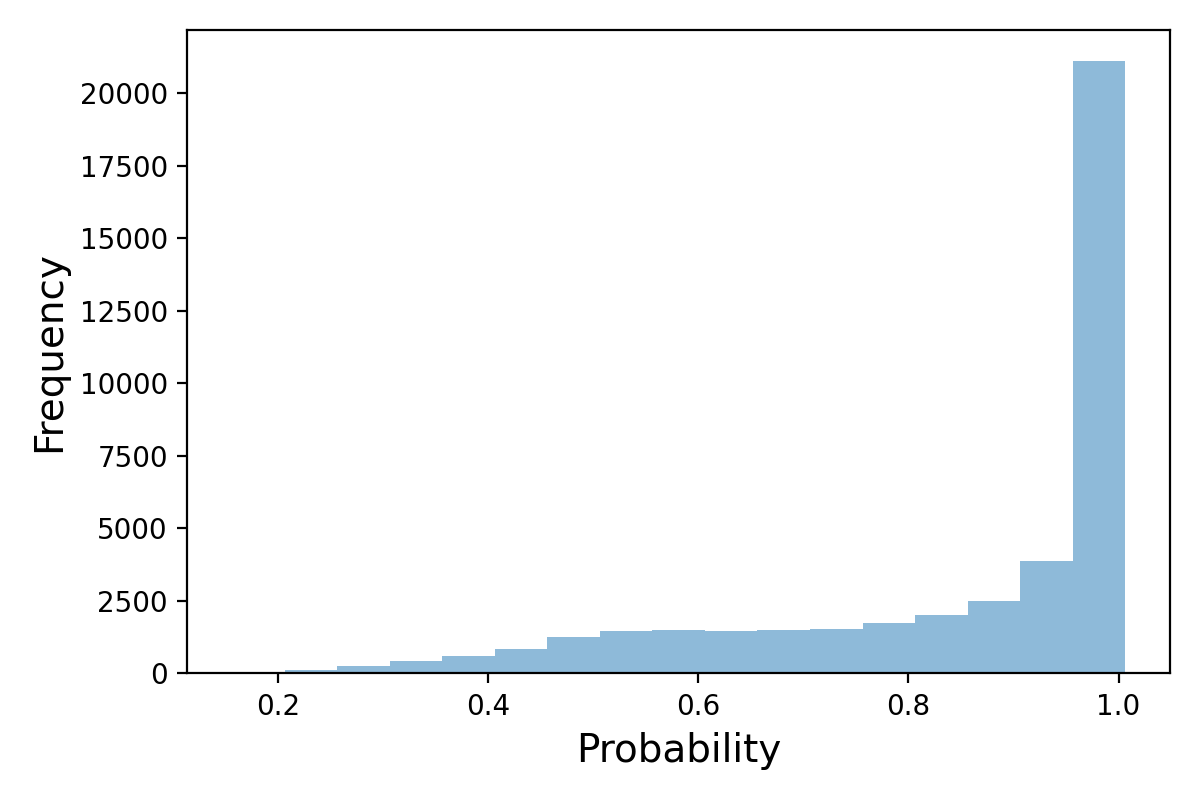}
\end{center}
\caption{Distribution of the deep ensemble class probabilities for all $\num{42230}$ images in the unlabelled sample, using a bin width of 0.05.}
\label{fig:probability_histogram}
\end{figure}

We display the distribution of class probabilities in Fig.~\ref{fig:probability_histogram}. These probabilities range from 0.16 to 1, with $60\%$ of the sources having a class probability greater than 0.9.

\begin{figure}
\begin{center}
\includegraphics[width=0.99\columnwidth]{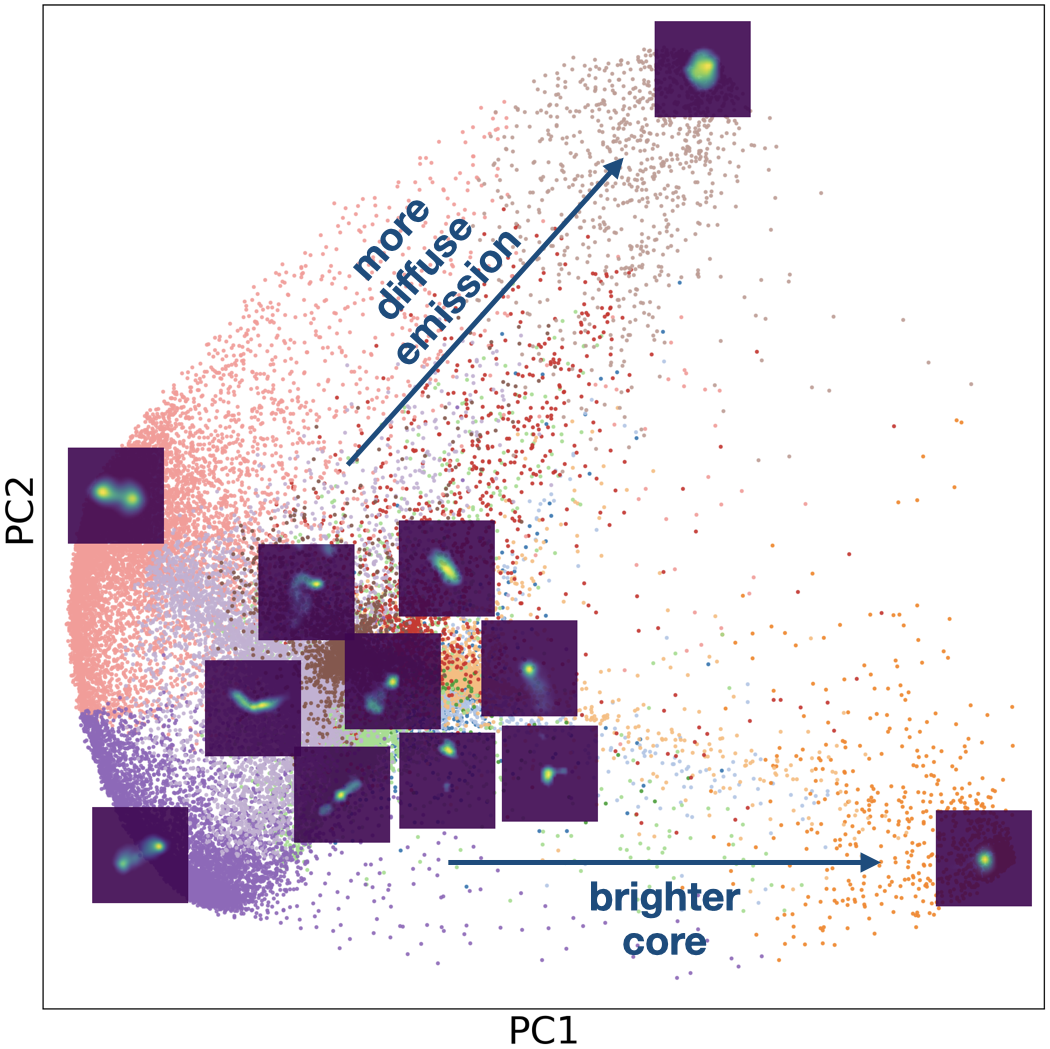}
\end{center}
\caption{Distribution of deep ensemble representations for the entire unlabelled sample in the space defined by the first two principal components, with colours indicating predicted classes. An example image for each class is included to illustrate the morphological evolution within this space.}
\label{fig:ensemble_distribution_images}
\end{figure}

Next, we analyse the distribution of these sources in the space defined by the first two principal components of the ensemble representations. This distribution is shown in Fig.~\ref{fig:ensemble_distribution_images}, where a prototypical image is provided for each class. The clusters containing bright core and circular diffuse sources are still the most distant ones (as observed with a single fine-tuned model). Meanwhile, the gap between double-component sources and the other clusters has ben partially filled by the previously unlabelled sources.

The distribution indicates that there are sources linking the main cluster block with the two more distant clusters. Visual inspection reveals that these sources exhibit morphologies that become more similar to those of the distant clusters as they draw nearer. For the sources approaching the bright core cluster, a bright circular structure begins to dominate their morphology. Meanwhile, sources moving closer to the circular diffuse cluster increasingly show a bright intensity peak surrounded by diffuse emission. We describe the morphological evolution in the distribution below.
\begin{itemize}
    \item The active pixel fraction increases in clockwise direction from the southeast to the northeast with the double-component sources located midway.
    \item At the centre of the distribution, we find the bent sources, which include both centre-bright and edge-bright sources.
    \item To their right, three core-dominated clusters form a diagonal. This starts as the lower cluster, which contains centre-bright sources with weak lobes. In the middle, there are single lobe sources with a single more diffuse lobe and the upper cluster consists of ellipsoidal centre-dominated sources.
    \item Below this diagonal, the artifacts and amorphous sources are situated very close to each other.
\end{itemize}

To evaluate the quality of the predicted classes and their associated probabilities, we manually labelled a subset of the unlabelled sample, including representatives from all 12 classes identified by the model. The class distribution of this manually labelled subset is presented in Table \ref{tab:manual_distribution}. Appendix \ref{sect:classification_evaluation} provides a comparison of the model-assigned labels with those determined through visual inspection. This comparison is based on F1 scores, with Table \ref{tab:classification_metrics} showing the F1 scores for each class and the macro-averaged score. We find that the F1 scores correlate with predicted class probabilities, yielding a macro-averaged F1 score of 0.60 for the entire visually inspected subset, 0.37 for images with predicted class probabilities below 85\%, and 0.80 for images with predicted class probability above 99\%. Despite expected confusion in the uncurated dataset, the classification provides a good foundation for analysing a new set of radio galaxies.

To quantitatively validate the intuition that the model distinguishes radio galaxy features such as centre brightness, edge brightness, and the bending angle, based on visual inspection of our results, we compare the model-assigned labels with the conventional classification scheme from \cite{2023DIB....4708974G}, which includes FRI, FRII, and bent classes, in Appendix \ref{sect:widespread_scheme}. Labels for this comparison are obtained through visual inspection and certain predicted classes are grouped into FRI, FRII, and bent superclasses to facilitate the comparison. We find an overall accuracy of the superclasses of $84\%$, with the main cause of confusion (see Fig.~\ref{fig:confusion_matrix}) being that the FRI or FRII properties are not independent of the bending angle of the source, confirming that the model distinguishes these radio galaxy features.

Additionally, we examine the source positions in the first two principal components of the fine-tuned representations with respect to class probability. We find that the 3291 sources with class probabilities below 0.5 (representing 7.8\% of the unlabelled sample) consistently lie between the class clusters, as shown in Fig.~\ref{fig:low_probability_sources}.

\section{Discussion}
\label{sect:discussion}
In this section, we assess the strengths and limitations of our methodology, followed by a comparative analysis of our approach with alternative methods.

\subsection{Evaluation of the methodology}
In this project, we explored how deep clustering with SSL can classify sources in radio astronomical surveys, focusing specifically on LoTSS-DR2. We found that pre-training an encoder using the learning task proposed in \cite{simclr} to predict representations that primarily capture morphological information of radio galaxies, beyond centre-bright and edge-bright examples, is very challenging. The random augmentations suggested in \cite{simclr} are insufficient for the model to distinguish the diverse morphological properties of radio galaxies. Consequently, we developed a new random augmentation, named RSV (see Sect.~\ref{sect:random_augmentation}), based on image-level similarity, to tailor the contrastive learning task to the radio astronomical context.

The random augmentation RSV uses SI-SSIM to measure source similarity at the image-level. This metric could be enhanced to be size-invariant, which would allow, for example, two symmetric double sources to have the same SI-SSIM value regardless of their apparent size in the image. Additionally, exploring alternative metrics could yield options that are potentially more suitable for this task. When applying the random augmentation to much larger datasets, such as those anticipated in future surveys, we  need to improve the computational efficiency of SI-SSIM matrix computation.

The introduction of this new augmentation allowed for the identification of additional, more complex morphological classes, notably bent sources, within the unlabelled sample. This approach operates under the assumption that the most similar sources, as determined by the similarity metric in image space, belong to the same morphological class. While this assumption has limitations and could potentially lead to false positive pairings that might counteract or bias the model's learning process, we did not observe significant issues such as unstable training or strongly biased classification results in our work. This limitation is further mitigated by using both augmented views from the same image with a $1/k$ probability, where $k$ is the number of SI-SSIM neighbours used in the augmentation. In future works, it would be beneficial to assess the likelihood of false positive pairings using a labelled dataset to enhance our understanding of this random augmentation.

We found that we were unable to fully handle the complete morphological heterogeneity present in the survey, particularly when considering all resolved sources in the catalogue. Consequently, we had to apply stringent cuts to the dataset, which both improved the clarity of source appearance in the images and slightly reduced the morphological diversity of the sources (see Sect.~\ref{sect:data}). After reducing the size of the unlabelled sample, we pre-trained the SSL model with randomly initialised model weights. For future works, it could be interesting to first carry out a pre-training on the whole data release and afterwards continue the pre-training on the filtered dataset. Furthermore, we do not rule out that a different method or additional random augmentations could accommodate the complete morphological diversity. Methodological modifications could include using vision transformers \citep{2020arXiv201011929D}, potentially with increased model size, in place of convolution-based encoders, as well as adopting a generative learning framework, such as the one proposed by \cite{2021arXiv210608254B}, instead of the invariance-based framework used in this work.

We would like to emphasise that the classification scheme used in the linear evaluation protocol can significantly influence the effectiveness of capturing morphological information within the representations. Therefore, it is advantageous to have a clean, large, labelled dataset that encompasses a diverse range of morphological classes.

The resulting representation space successfully incorporates the morphological information we aimed to capture. Among the morphological properties encoded in the representation vectors are data quality (including artefacts and amorphous sources), source type (e.g. radio halo, bright core, or radio galaxy), the number of source components, the relative brightness of core and hotspots, and the degree of source bending. Some properties reflected in the representations, which are not intrinsically related to the source morphology, include the quality of observational calibration, source size, and source resolution.

\cite{simclr} reported the emergence of distinct clusters for different ImageNet \citep{imagenet} classes within the representation space (see Fig. B.4 in their original paper). Therefore, we anticipated achieving a similar representation space structure for the morphological classes of radio galaxy. Under this assumption, applying a clustering algorithm to the representations should facilitate the identification of radio galaxy classes, as suggested by \cite{2024MNRAS.tmp..951M}. However, the obtained representation space structure is more diffuse than required for that approach (see Fig.~\ref{fig:clusters_ssl}). This necessitated the creation of a labelled subsample tailored to the morphological diversity of the unlabelled sample to fine-tune the encoder and perform the classification as a downstream task (see Sect.~\ref{sect:finetuning}). We acknowledge that a more suitable model and a better-designed learning task could potentially lead to the desired emergence of morphological clusters.

The most significant difference between SimCLR \citep{simclr} and our work is likely the imbalanced nature of our dataset and the absence of class labels. It is known that SimCLR, along with other SSL algorithms, performs poorly when trained on imbalanced data due to its hidden uniform prior \citep{2022arXiv220407141A}. For future works, we propose using models that do not have this limitation, such as the extended version of the Masked Siamese Networks model presented in \cite{2022arXiv220407141A}.

Avoiding the use of class labels at the initial stage is a crucial aspect of our approach. However, the availability of class labels representing the entire class diversity of the dataset makes a significant difference, as it allows for a comprehensive evaluation to determine whether the information contained in the representations sufficiently describes the whole dataset with the linear evaluation protocol. This evaluation allows for a more efficient design of the learning problem,  for instance, by developing new random augmentations or combining loss functions.

Additionally, objects in natural images are often more distinctly quantised compared to those in radio sources. For instance, while a cat and an aeroplane may share some features, they are categorically distinct objects. In contrast, a centre-bright bent source and a linear edge-bright source are variations of the same object, influenced by different environmental conditions. Furthermore, the appearance of the radio galaxies in images is influenced by their redshift and the resolution achieved, posing a significant challenge when attempting to map these sources into an abstract representation space.

\cite{mingo} created a dataset of FRI and FRII radio galaxies using an automated algorithm complemented by visual analysis from the LoTSS-DR1 \citep{2019A&A...622A...1S}. They identified 1256 FRI and 423 FRII galaxies, with a subsample of 459 FRI sources exhibiting a bent morphology. Interestingly, our sample has fewer FRI sources than FRII sources with a class probability higher than 90\%. Specifically, we find 2682 FRI and 3774 FRII sources. Additionally, we found 4872 sources exhibiting bent morphology, which are classified as different categories in our dataset, in contrast to the classification approach used by \cite{mingo}. The differing class definitions between our work and theirs may explain the discrepancy in the dominant FR type observed in our samples. Another factor contributing to the discrepancy is the different selection functions employed in each study, which can result in significantly diverging populations. Our findings suggest the importance of exploring classifications beyond the traditional FRI and FRII categories to fully capture the diversity of radio sources from sky surveys.

By comparing the model-predicted labels with those obtained through visual inspection for a manually labelled subset, we achieve a macro-averaged F1 score of 0.60 and observe a correlation between the F1 scores and the model's confidence in its labels. Additionally, we evaluated the model's ability to identify radio galaxy features, such as centre brightness, edge brightness, and the bending angle, using the conventional classification scheme of FRI, FRII, and bent radio galaxies. To facilitate this comparison, we grouped certain predicted classes into these superclasses, achieving an overall accuracy of 84\% across the FRI, FRII, and bent classes, with FRII sources being classified nearly perfectly. The primary source of confusion arises from the non-exclusive nature of FR and bent properties. To address this, we propose exploring hierarchical clustering approaches in future work.

We would like to discuss potential applications of this work. Firstly, interesting clusters could serve as the basis for astrophysical analyses. Anomalies or sources with uncommon morphologies can be identified among images with the lowest class probabilities. Once an intriguing object is identified, the representation space from the pre-training phase could be used to search for similar objects, potentially uncovering a missing class. With a subset of approximately 30 sources exhibiting this morphology, this new `cluster' could be used to fine-tune the encoder and reclassify the dataset to identify all instances of this morphology. This process could be performed iteratively, leading to comprehensive and efficient morphological classification and representation space.
In the next step, the morphological classes that we found in this work will be plotted against other observables from the value added catalogue, such as stellar masses, or derived quantities such as galaxy overdensity to study the dependence on the environment.
Finally, the ability to explicitly incorporate complex morphological properties into the representation space while encompassing a wide variety of radio source morphologies is a significant step toward developing a foundation model for radio surveys. 

\subsection{Comparison to other methods}
In this work, we used deep clustering with SSL methods to study the morphological classification of radio galaxies that include a broader range of classes than those typically found in existing labelled datasets. Related works are the studies of SSL methods for radio astronomy and unsupervised classification.

Studies of SSL methods for radio astronomy include: The SSL pre-training with large amounts of unlabelled radio source images and the fine-tuning with FRI and FRII labelled images showing an improvement in classification accuracy and the advantage of reducing the number of required labelled images \citep{2023PrCS..222..601H, 2024RASTI...3...19S}. In our work, we use a similar SSL model as done in these studies and take advantage of the small number of labelled data required for fine-tuning this models.

\cite{2024MNRAS.tmp..951M} tested the unsupervised classification of FRI and FRII radio galaxies by combining an SSL model and clustering algorithm. We have tested this approach for our classification. However, the representation space structure is not clustered enough to obtain reasonable results with numerous classes.

Another way of classifying radio sources is by number of radio components and number of intensity peaks, as done in the RGZ with the help of citizen scientists. \cite{2024PASA...41...85R} also studied the use of an SSL algorithm to perform this task. Separately, the transfer of these labels with the help of an SOM and a random forest \citep{breiman2001random} was studied in \cite{2019PASP..131j8009G}. 

\cite{2024arXiv241114078C} performed a benchmark study comparing different SSL models and different classification schemes providing the linear evaluation for the different combinations. In contrast, we are unable to provide classification accuracies because we do not have true labels for the classified dataset.

The unsupervised classification of radio sources by their visible morphology was tested by \cite{2019PASP..131j8011R} using the combination of an autoencoder \citep{SANGER1989459}, a SOM, and the clustering algorithm $k$-means \citep{lloyd1982least}. Later, \cite{2021A&A...645A..89M} performed the same task using a rotation and flipping invariant SOM followed by manual grouping and labelling of the identified morphological classes. The latter was done within the search of the rarest morphologies in LoTSS-DR1. 

A great advantage of SOMs over SSL algorithms is their interpretability. We attempted to identify the morphological features detected by the SSL algorithm in each cluster through visual inspection. In contrast, SOMs learn prototypes that are optimised to best fit the training dataset. These prototypes show directly which features are  taken into account by the algorithm. The downside of SOMs is that they describe the dataset using a discrete number of prototypes, which must be selected empirically. Meanwhile, our algorithm first learns a continuous representation of the input data, having more degrees of freedom to describe the dataset. Additionally, the use of representations has the advantage of enabling the use of other data types, such as time-series data, spectra, or language data; for instance, verbal descriptions of the morphology \citep{2023MNRAS.522.2584B}.

 In addition to the radio frequency maps, \cite{2019PASP..131j8009G} used infrared observations. This is an essential improvement, since infrared or optical maps enable the identification of the host galaxy location. This helps to determine the associated components and improves the quality of the morphological class estimation. This could also be implement in the SSL approach, by adding these maps as an additional channel to the input images; for instance, by extracting the same sky region from the corresponding survey. \cite{2020MNRAS.497.2730G} showed with this approach that a SOM could group radio components and identify the host galaxy.

\section{Conclusions}
\label{sect:conclusions}

In this work, we investigated the morphological classification of a subset of resolved sources from LoTSS-DR2 using self-supervised deep clustering. The final classification consists of 12 classes, including known classes of radio galaxies, such as bent-tail sources, as well as new classes. Our classification scheme incorporates distinctions among core-bright, edge-bright and bent sources (also used in previous classification studies) by considering the overall morphological shape and symmetry of the radio emission.\\

The characteristic feature of our approach is the creation of a small labelled subsample using the representation space learned during pre-training. This strategy aims to capture the morphological diversity of the unlabelled sample within the identified classes as exhaustively as possible, thereby improving the significance of the final source classification.
To create the labelled subsample, we required a representation space that encompassed extensive morphological information. To enhance the morphological information within the representations, we designed a random augmentation, RSV. This augmentation uses the most similar images, evaluated at the image level with a modified version of SSIM to create the second view for the SSL learning task.

Our training strategy involved pre-training using SSL with all images contained in the unlabelled sample, fine-tuning the deep ensemble with the created labelled subsample, and then predicting class probabilities for all images in the unlabelled sample to achieve the final classification.

The deep ensemble can be used to infer the class probabilities of radio source images unseen by the model. We also study the meaning of the position in the fine-tuned representation space. Additionally, the pre-trained model can be used to search for images that are morphologically similar to a query image within its neighbouring regions in the representation space.

The identification of the single lobe class in the unlabelled sample suggests that our technique has the potential to uncover morphological classes that may not be readily apparent to the eye. Finally, the improvement of the learned representations regarding morphological information represents a step towards the unsupervised classification of enormous astronomical surveys and the development of foundation models.

\cite{2023MNRAS.520.1439L} used active learning as an extension of the pipeline to successfully identify anomalies, leading to the discovery of a ring-like radio source in the MeerKAT Galaxy Cluster Legacy Survey.
Obtaining high-quality representations could enable the identification of anomalies, namely, sources with uncommon morphologies, in radio survey data with SSL algorithms.

\begin{acknowledgements}
 
We express our gratitude to the anonymous referee for their insightful and thorough report. Funded by the Deutsche Forschungsgemeinschaft (DFG, German Research Foundation) – project number 460248186 (PUNCH4NFDI).
GK, MB, and LLS acknowledge funding by the DFG under Germany's Excellence Strategy -- EXC 2121 ``Quantum Universe'' --  390833306.

\end{acknowledgements}

\bibliographystyle{aa}
\bibliography{references}

\begin{appendix}

\section{Random structural view ablation study}
\label{sect:ablation}

We compare SimCLR pre-training with and without the RSV augmentation in the following. The goal of introducing RSV is to improve clustering in the representation space, so that sources with common morphology have nearly identical representation vectors. Therefore, we expect that improved clustering will lead to a deterioration in the metrics that measure the identification of one augmented view given the other (i.e., the loss function and accuracies), due to confusion caused by other sources with the same morphology having very similar representation vectors.

Table \ref{tab:srv_ablation} presents the loss and evaluation metric values (refer to Sect.~\ref{sect:ssl_metrics}) for both models. We observe that SimCLR without RSV nearly achieves perfect values for the NT-Xent loss and accuracy metrics, with a top-1 accuracy of 1.00 and a mean position accuracy close to 1. This indicates that the model can perfectly distinguish between two augmented views of a given input image and all other augmented views within the batch, suggesting that the model is focusing on features not explicitly related to the morphology of radio galaxies. In contrast, SimCLR trained with RSV has significantly lower values, with a top-1 accuracy of 0.19 and a mean position accuracy of 49. Taking into account the difference in the SNN metric of 3.58 between the models, this suggests that the model trained with RSV is concentrating on features more related to morphology, making it more challenging to distinguish sources with similar morphologies. Additionally, this indicates that RSV enhances morphological clustering in the representation space, causing the most similar images, as identified by SI-SSIM, to be significantly more cohesive than other images in the batch. Finally, we find that the representations from both models, when used in a linear model, effectively differentiate between linear and bent centre-bright and edge-bright sources.

\begin{table}[h!]
\caption{Performance comparison of SimCLR pre-training with and without RSV.}
\label{tab:srv_ablation}
\centering
{
\renewcommand{\arraystretch}{1.3}
\begin{tabularx}{\columnwidth}{l c c}
\hline
\textbf{Loss + metric} & \textbf{with RSV} & \textbf{without RSV} \\
\hline
NT-Xent & 4.48 & 0.08 \\
top-1 accuracy & 0.19 & 1.00 \\
mean position accuracy & 49.00 & 1.06 \\
SNN & 1.68 & 5.26 \\
linear evaluation & 0.90 / 0.89 / 0.91 & 0.81 / 0.94 / 0.93
\end{tabularx}
}
\tablefoot{Pre-training loss and evaluation metrics are presented for both models. For the linear evaluation protocol, values are provided separately for the training, validation and test sets.}
\end{table}
\FloatBarrier

Next, we examine how these metrics translate into the morphological classes we can recover within the pre-labels. After conducting the clustering scan (see Sect.~\ref{sect:clustering}), we identify 95 clusters with the model including RSV and 82 clusters without RSV. Although the number of clusters is relatively similar, the model trained without RSV fails to capture the morphological classes: bright core, single lobe, diffuse bent and structured bent. These classes represent one-third of the total number of classes identified in the unlabelled sample. Therefore, we conclude this section by acknowledging that the inclusion of the random augmentation enables the model to detect additional morphologies, in particular bent sources.

\section{Classification evaluation}
\label{sect:classification_evaluation}

In this section, we compare the class labels assigned by our algorithm with those determined through visual inspection. To this end, we manually labelled 1169 images, covering all 12 classes identified by our method. Manual labelling proved challenging due to the model's unknown classification criteria, inherent uncertainties in human judgement, and the rarity of certain morphologies. Consequently, the dataset is highly imbalanced, with the class distribution presented in Table \ref{tab:manual_distribution}. Additionally, we provide the counts for subsets of images with predicted class probability below 85\% and above 99\%.

\begin{table}[h!]
\caption{Class distribution of manually labelled images.}
\label{tab:manual_distribution}
\centering
{
\renewcommand{\arraystretch}{1.3}
\begin{tabularx}{\columnwidth}{l c c c}
\hline
\textbf{Class name} & $\mathbf{p\leq 85\%}$ & \textbf{Entire sub-set} & $\mathbf{p\geq 99\%}$ \\
\hline
artifact                       & 5 & 20 & 10 \\
amorphous                      & 22 & 66 & 26 \\
bright core                    & 13 & 62 & 34 \\
head-tail                      & 4 & 15 & 5 \\
single lobe                    & 2 & 9 & 6 \\
centre-bright                  & 26 & 102 & 44 \\
c. p. ellipse                  & 12 & 32 & 13 \\
symmetric double               & 52 & 211 & 116 \\
edge-bright                    & 108 & 327 & 85 \\
diffuse bent                   & 76 & 167 & 40 \\
structured bent                & 29 & 93 & 34 \\
circular diffuse               & 18 & 65 & 33 \\
\hline
Total & 367 & 1169 & 446
\end{tabularx}
}
\tablefoot{The counts are provided for the entire subset, the subset with predicted class probabilities below 85\%, and the subset with predicted class probabilities above 99\%.}
\end{table}
\FloatBarrier

Table \ref{tab:classification_metrics} presents the F1 scores for each class, with values provided for the entire subset and for the subsets with predicted class probabilities below 85\% and above 99\%. We observe considerable variation in scores across classes, with the centre-bright class achieving the highest score of 0.73 and the head-tail class the lowest score of 0.36 for the entire subset. The macro average, where each class is given equal weight, yields an overall score of 0.60 for the entire subset. We find that the macro-average score decreases to 0.37 for images with predicted class probabilities below 85\% and increases to 0.80 for images with predicted class probabilities above 99\%. Furthermore, the ranking of the classes based on these scores varies across subsets with different confidence levels.

\begin{table}[h!]
\caption{F1 scores for each class and the corresponding macro average.}
\label{tab:classification_metrics}
\centering
{
\renewcommand{\arraystretch}{1.3}
\begin{tabularx}{\columnwidth}{l c c c}
\hline
\textbf{Class} & $\mathbf{p\leq 85\%}$ & \textbf{Entire sub-set} & $\mathbf{p\geq 99\%}$ \\
\hline
artifact                       & 0.22 & 0.59 & 0.77 \\
amorphous                      & 0.31 & 0.67 & 0.92 \\
bright core                    & 0.00 & 0.63 & 0.85 \\
head-tail                      & 0.17 & 0.36 & 0.59 \\
single lobe                    & 0.22 & 0.48 & 0.83 \\
centre-bright                  & 0.40 & 0.73 & 0.91 \\
c. p. ellipse                  & 0.33 & 0.42 & 0.70 \\
symmetric double               & 0.41 & 0.64 & 0.87 \\
edge-bright                    & 0.54 & 0.65 & 0.81 \\
diffuse bent                   & 0.40 & 0.47 & 0.59 \\
structured bent                & 0.26 & 0.51 & 0.67 \\
circular diffuse               & 0.53 & 0.66 & 0.80 \\
\hline
\textbf{Macro average} & \textbf{0.37} & \textbf{0.60} & \textbf{0.80}
\end{tabularx}
}
\tablefoot{The values are compared across the entire subset, the subset with predicted class probabilities below 85\%, and the subset with predicted class probabilities above 99\%.}
\end{table}
\FloatBarrier

Although the evaluation is not ideal, given the fundamentally unlabelled nature of the dataset, it still provides some valuable insights. Firstly, as expected from a model handling uncurated data, there is considerable confusion among the classes. Secondly, the extend of this confusion correlates with the class probabilities predicted by the model. Thirdly, the metric values support the impressions gained from visually inspecting the classification results - namely, that the classification can serve as an initial separation of a dataset, providing a useful starting point for more detail analysis.

\section{Widespread classification scheme}
\label{sect:widespread_scheme}

In the following, we compare the classification scheme identified by our model with the widely used scheme of FRI, FRII and bent sources. This comparison enables a quantitative evaluation of the model's ability to identify the radio galaxy features, including centre-bright, edge-bright and bending angle characteristics.

Due to the specific range of source sizes, there is minimal overlap between the available labelled radio galaxy samples, such as the set compiled in \cite{2023DIB....4708974G}, and our unlabelled sample. Consequently, we labelled a subset of 300 sources through visual inspection and employed the conventional classification scheme from \cite{2023DIB....4708974G}. This subset includes 100 sources each of FRI, FRII and bent sources, thus constituting a class-balanced set. For comparing both classification schemes, we group certain predicted classes into superclasses: the centre-bright and centrally peaked ellipse classes are associated with the FRI morphology, the symmetric double and edge-bright classes with the FRII morphology, and the diffuse bent and structured bent classes with the bent morphology.

\begin{figure}[h!]
\begin{center}
\includegraphics[width=0.8\columnwidth]{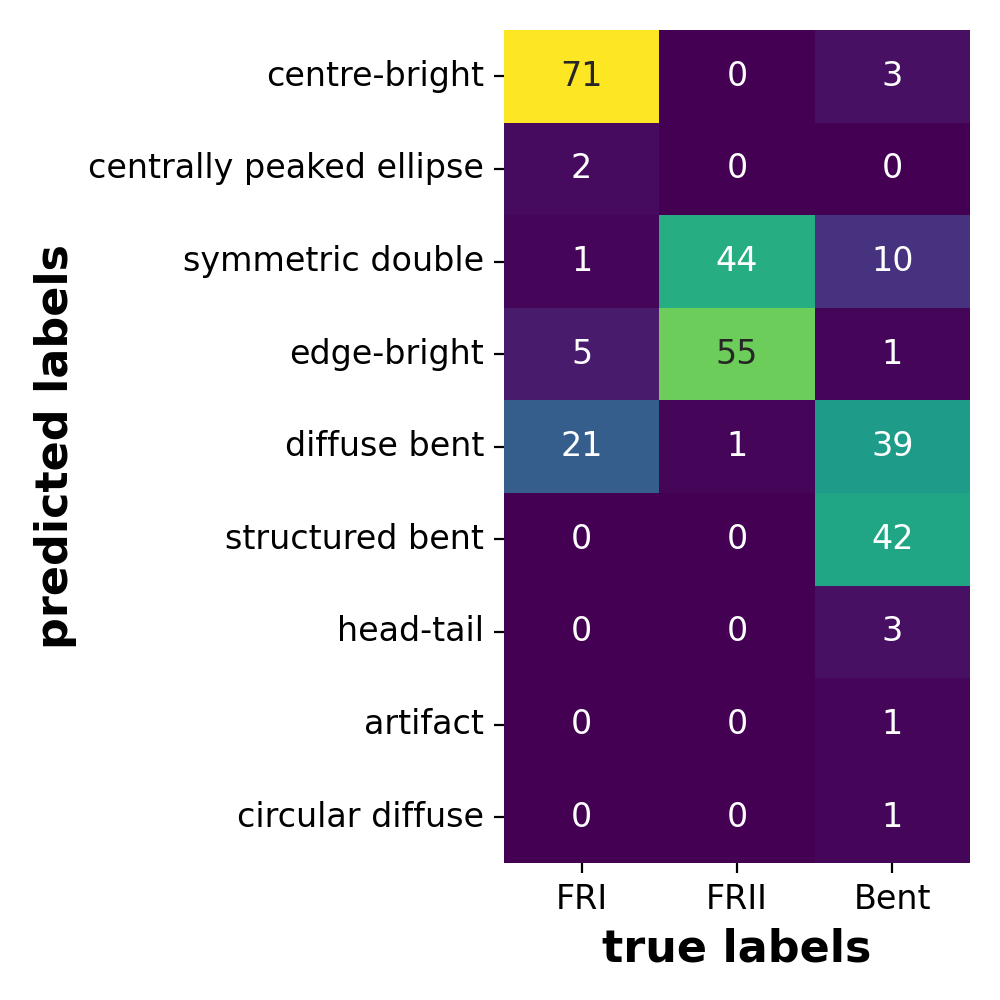}
\end{center}
\caption{Confusion matrix comparing the classes predicted by our model to the conventional classes determined through visual inspection. The first two predicted labels correspond to FRI sources, the next pair to FRII sources, and the subsequent pair to bent ones. The remaining predicted labels represent other classes.}
\label{fig:confusion_matrix}
\end{figure}
\FloatBarrier

\begin{figure}[h!]
\begin{center}
\includegraphics[width=0.99\columnwidth]{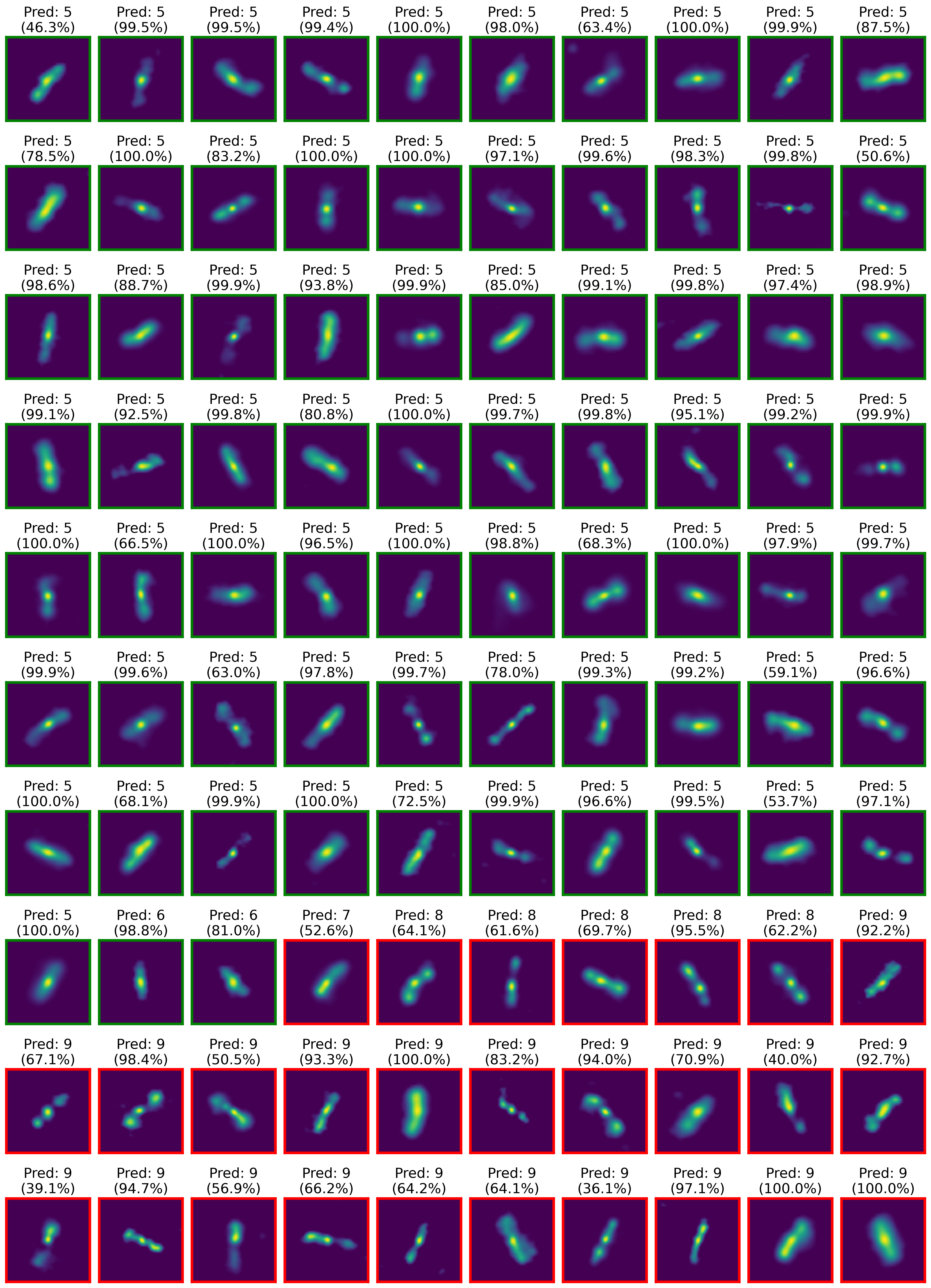}
\end{center}
\caption{Image grid displaying the 100 sources manually labelled as FRI, accompanied by the class and class probability predicted by our model. Sources with predicted labels compatible with FRI (centre-bright or centrally peaked ellipse) are framed in green, whereas those with different predicted labels are framed in red. The predicted class number corresponds to those shown in Table~\ref{tab:class_population}.}
\label{fig:fri_labelled}
\end{figure}

\begin{figure}[h!]
\begin{center}
\includegraphics[width=0.99\columnwidth]{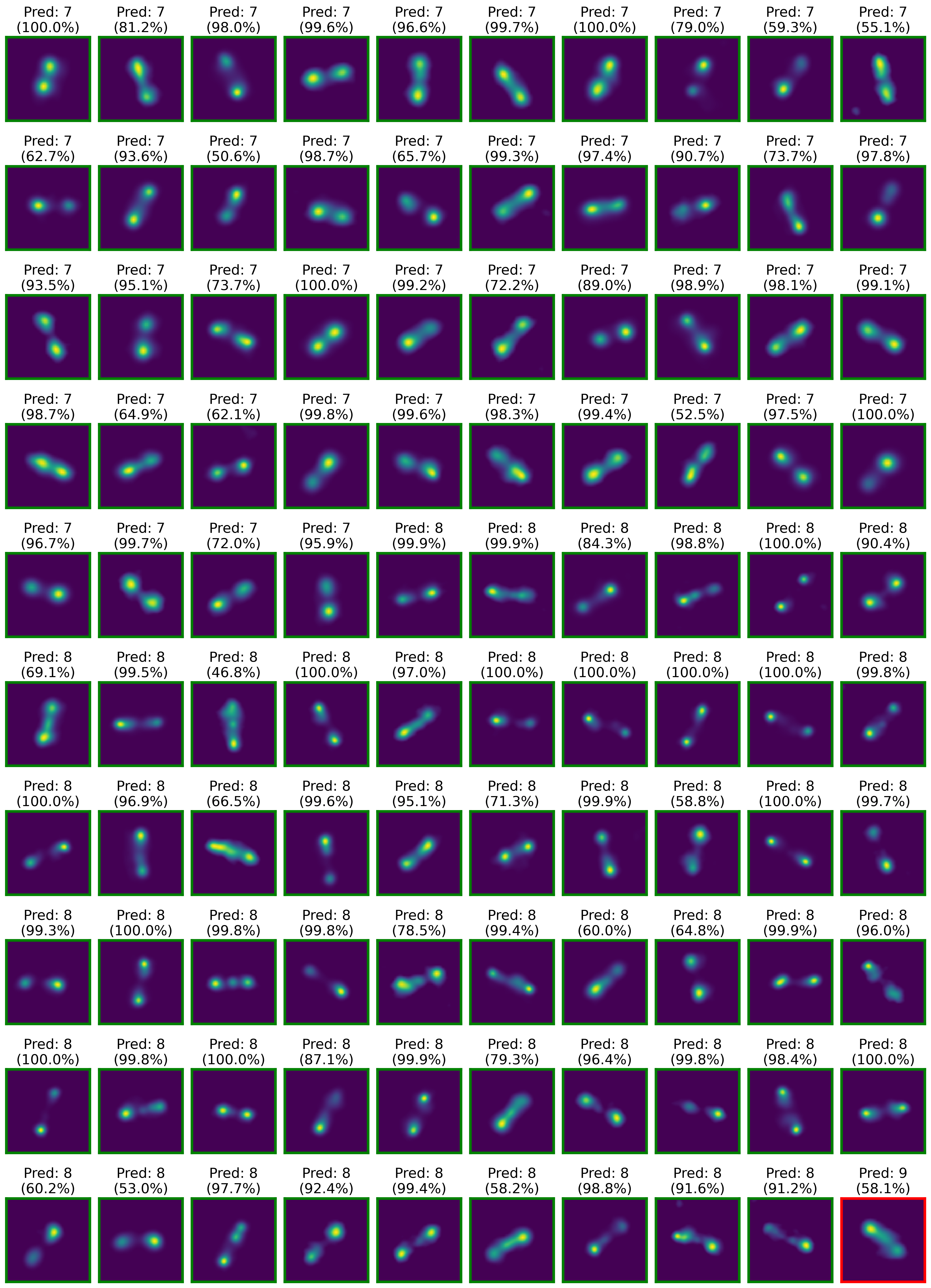}
\end{center}
\caption{Image grid displaying the 100 sources manually labelled as FRII, accompanied by the class and class probability predicted by our model. Sources with predicted labels compatible with FRII (symmetric double or edge-bright) are framed in green, whereas those with different predicted labels are framed in red.}
\label{fig:frii_labelled}
\end{figure}

\begin{figure}[h!]
\begin{center}
\includegraphics[width=0.99\columnwidth]{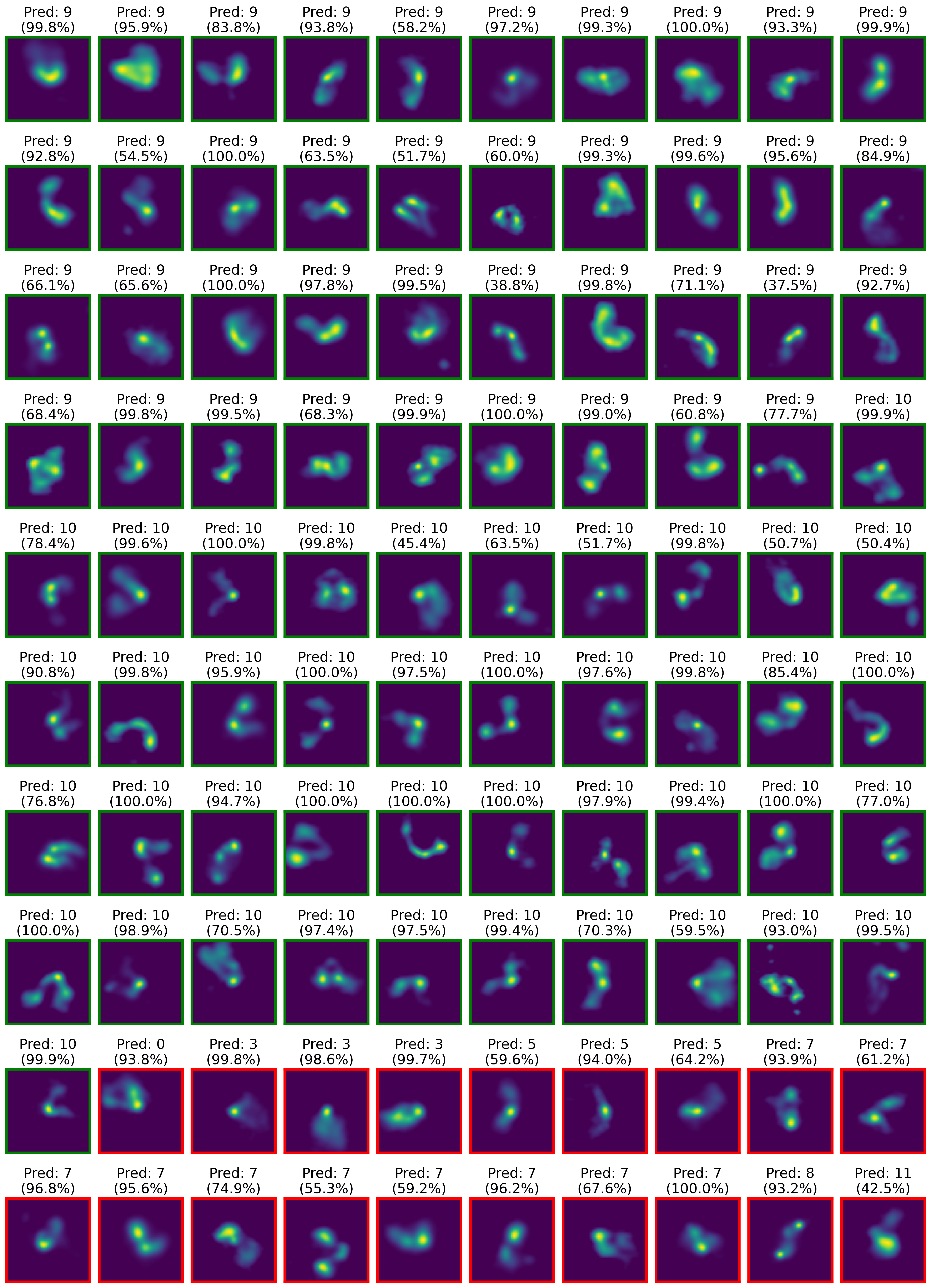}
\end{center}
\caption{Image grid displaying the 100 sources manually labelled as bent, accompanied by the class and class probability predicted by our model. Sources with predicted labels compatible with bent (diffuse bent or structured bent) are framed in green, whereas those with different predicted labels are framed in red.}
\label{fig:bent_labelled}
\end{figure}

The comparison of labels yields accuracies of 73\%, 99\%, and 81\% for FRI, FRII, and bent sources, respectively, resulting in an overall mean accuracy of 84\%. Our findings indicate that FRII sources are nearly perfectly  distinguished, while FRI and bent sources exhibit some degree of confusion. To further elucidate these results, we present the confusion matrix in Fig.~\ref{fig:confusion_matrix}, which illustrates the relationship between our predicted classes and the manually labelled ones. Most FRI-labelled sources are correctly classified as centre-bright sources, although a notable portion is classified as diffuse bent. This mixing, while undesired, is not surprising given that a single source can exhibit both centre-bright and bent characteristics. To better understand the mixing, we present Fig.~\ref{fig:fri_labelled} showing the FRI-labelled sources along with their predicted class and class probability. We find that sources predicted to be bent with high probability also present some bending. Regarding the FRII-labelled sources, only one is misclassified as diffuse bent, which can again be explained by the non-exclusive nature of these properties. By inspecting the sources manually labelled as FRII with their corresponding predicted class and class probability in Fig.~\ref{fig:frii_labelled}, we find that the misclassified source indeed exhibits a small bending angle. Finally, bent sources are predominantly classified as diffuse bent and structured bent, with the next most predicted class being the incorrectly identified symmetric double. In Fig.~\ref{fig:bent_labelled}, we observe that the sources predicted to be symmetric doubles with high probability exhibit the characteristic symmetric double structure, along with additional, less dominant radiation that the model seems to overlook.

\FloatBarrier

\FloatBarrier

\FloatBarrier

\section{Representation distribution and class probability}
\label{sect:representation_dist}

\begin{figure}[h!]
\begin{center}
\includegraphics[width=0.9\columnwidth]{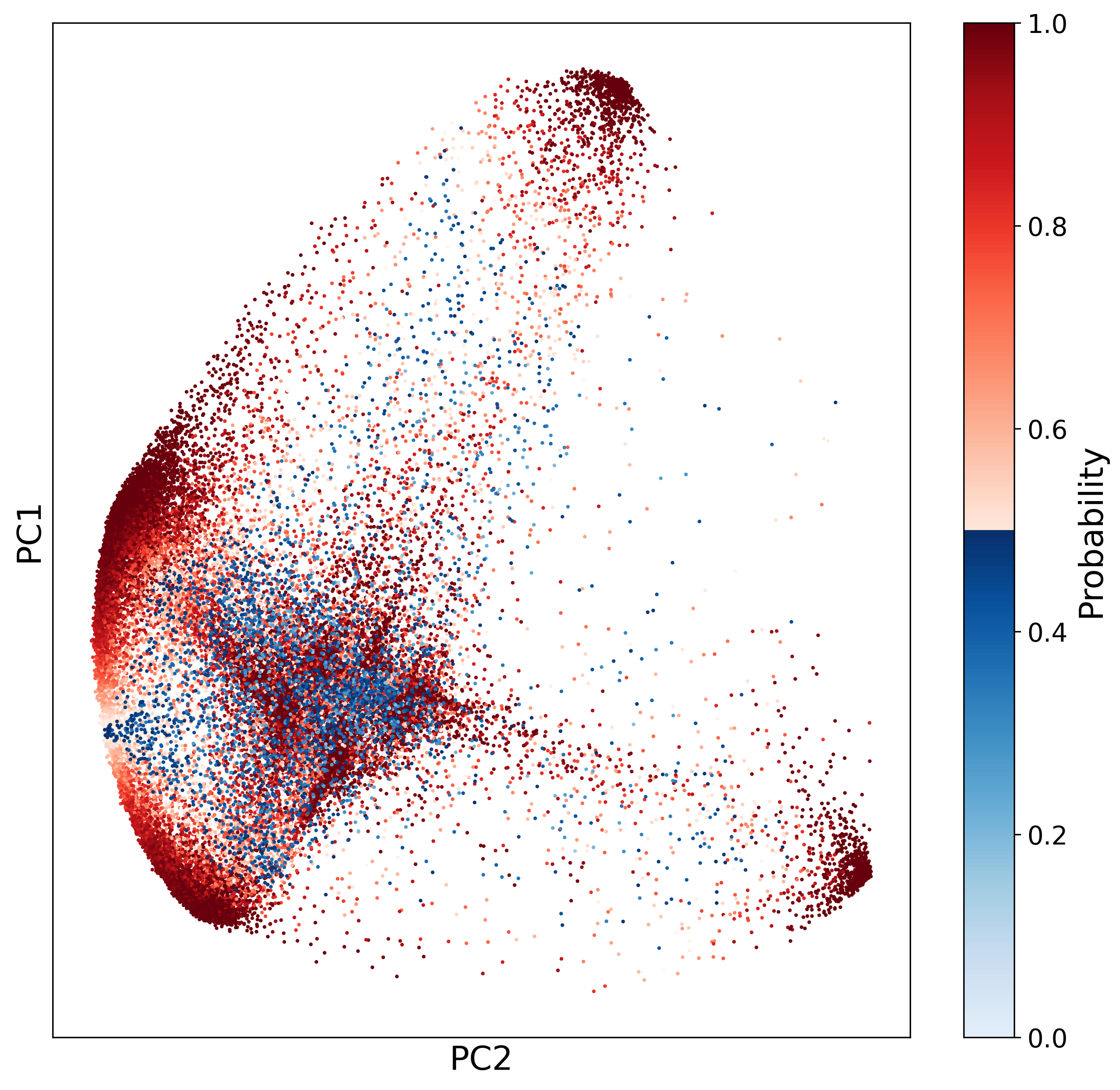}
\end{center}
\caption{Distribution of deep ensemble representations projected onto the space defined by the first two principal components. Data points are coloured blue if their class probability is less than 0.5 and red if their class probability is 0.5 or greater.}
\label{fig:low_probability_sources}
\end{figure}

To study the relation between predicted class probability and position in representation space, we display the distribution of the first two principal components of the fine-tuned representations in Fig.~\ref{fig:low_probability_sources} distinguishing the sources with class probability less than 0.5. For better visualisation, we colour these points in blue and the rest in red. We find that the 3291 sources with class probability less than 0.5 are situated between the class clusters formed in the fine-tuned representation space.

\FloatBarrier

\end{appendix}

\end{document}